\documentclass[12pt,english,showpacs,showkeys]{revtex4}
\usepackage{multirow}
\usepackage{graphicx}
\usepackage{amssymb}
\usepackage{color}
\usepackage{amsmath}
\usepackage{caption}
\usepackage{ulem}

\makeatletter

\usepackage{babel}
\makeatother

\begin{document}

\title{DNA viewed as an out-of-equilibrium structure}

\author{
A. Provata$^{1,3}$\footnote{E-mail: aprovata@chem.demokritos.gr}, 
C. Nicolis$^2$\footnote{E-mail: cnicolis@oma.be},  
and G. Nicolis$^{3}$\footnote{E-mail: gnicolis@ulb.ac.be}
}

\affiliation{$^1$Department of Physical Chemistry, National Center for Scientific
Research ``Demokritos\char`\"{}, 15310 Athens, Greece \\
$^2$Institut Royal Meteorologique de Belgique, 3 Avenue Circulaire,
1180 Bruxelles, Belgium\\
$^3$ Interdisciplinary Center for Nonlinear Phenomena and Complex Systems,
Universit\'e Libre de Bruxelles, Campus Plaine, CP. 231, 1050 Bruxelles, Belgium}

\begin{abstract}
\textbf {Abstract:} The complexity of the primary structure of human DNA is explored using
methods from nonequilibrium statistical mechanics, dynamical systems theory and 
information theory. The use of chi-square tests shows that DNA cannot be described as
a low order Markov chain of order up to $r=6$. 
 Although detailed balance seems to hold at the level of
purine-pyrimidine notation it fails when all four basepairs are considered,
 suggesting spatial asymmetry and irreversibility. Furthermore, the block entropy does not
increase linearly with the block size, reflecting the
long range nature of the correlations in the human genomic sequences. To probe locally
 the spatial structure of the chain we study the exit distances from a specific symbol,
the distribution of recurrence distances and the Hurst exponent, all of which show power law tails
and long range characteristics. These results suggest that
human DNA can be viewed as a non-equilibrium structure maintained in its state
through interactions with a constantly changing environment. 
Based solely on the  exit distance
distribution accounting for the nonequilibrium statistics
and using the Monte Carlo rejection sampling method we construct a model DNA sequence.
This method allows to keep all long range and short range statistical characteristics
of the original sequence. The model sequence presents the same characteristic exponents
as the natural DNA but fails to capture point-to-point details.
 
\end{abstract}


\date{\today}

\maketitle

\section{Introduction}
\label{sec:intro}
The DNA molecule is one of the most complex systems encountered in nature. By its intricate, 
aperiodic structure it constitutes an information source for the synthesis of the different 
entities and for the occurrence of the multitude of delicately balanced processes within living
 cells. Yet the connection between global DNA structure and its various functions remains to a 
large extent elusive, particularly in view of the coexistence of coding and non-coding regions 
and the realization of the important role of non-coding sequences in higher organisms
\cite{alberts:2007,dunham:2012,arneodo:2011}. 

One view of the DNA molecule frequently adopted in the literature is that of two nested 
non-overlapping symbolic sequences - the coding and non-coding regions - each of which is expressed 
in terms of four symbols/letters corresponding to the four bases A, C, G and T or in the 
contracted form of two symbols/letters, purines (A, G) and pyrimidines (T, C). 
The observed complexity of these nested sequences has been shaped during 
evolutionary time based on functional needs. Processes like 
single nucleotide mutations, insertion and deletion
of segments, multiple repetitions of elements acting simultaneously over different length
and time scales have shaped the complexity of current day genomes producing intriguing 
statistical properties 
\cite{arneodo:2011,roman:1998,messer:2007,carpena:2011,polak:2009,oliver:2008}. 
In this latter setting early investigations have shown that the succession of bases along coding regions
in higher organisms presents short range correlations, whereas non-coding 
regions exhibit long-range correlations 
\cite{li:1992,stanley:1992,voss:1992}. 
 For these organisms
 the coding segment length distribution has an exponentially falling tail whereas the 
non-coding segment one falls off as a power law \cite{almirantis:2001,oikonomou:2007}.

In the present work the structure of DNA, viewed as a symbolic sequence, is analyzed from 
the standpoint of nonequilibrium statistical mechanics, dynamical systems theory and 
information theory. A first question raised concerns spatial asymmetry along the sequence, its 
signatures and its role in information processing. A second question pertains to the 
identification and analysis of global indicators of the underlying complexity, beyond 
the linear correlations usually considered in the literature. To arrive at a quantitative 
formulation of these questions we view a DNA chain as the realization of a stationary 
stochastic process, i.e., a process where the probabilities $p_i$ of the different 
states/symbols attain rapidly limiting values as the sample size is increased, in which 
the role of the time step in the traditional setting is played by the “spatial” shift 
by one unit in sequence space. 

The data along with the results of a preliminary statistical analysis are compiled in Sec. 
\ref{sec:data}. Section \ref{sec:markov} is devoted to a Markov chain analysis, leading to 
the conclusion that the data cannot be fitted by a Markov chain of order up to 6. The issue 
of spatial asymmetry is addressed in Sec. \ref{sec:fluxes} in which probability fluxes are 
evaluated and shown to be different from zero, reflecting the breakdown of (generalized) 
detailed balance type conditions. In Sec. 
\ref{sec:entropy} this analysis is complemented by the evaluation of a series of entropy 
and information-like quantities, leading to interesting characterizations of the “dynamical” 
complexity as one advances along the original chain and along its reverse and of the 
information transfer between different parts of the chain. Exit distance and recurrence 
distance distributions, two global complexity indicators of special significance are computed 
from the data and analyzed in Sec. 
\ref{sec:exit-distances}. The existence of long tails in the distributions and of long-range 
correlations in the associated lengths is established and confirmed further by the evaluation 
of Hurst exponents. Building on this information a construction algorithm of a “model DNA” 
possessing the same statistical properties as the natural one, free of extra assumptions, 
is outlined in Sec. \ref{sec:model}.
 Different criteria for comparing model and natural DNA’s are also developed. The main 
conclusions are summarized in Sec. \ref{sec:concl}. 

\section{Data and statistical analysis}
\label{sec:data}

\par 
For the needs of our analysis we have employed the genomic data
from two large human chromosomes (10 and 14) and two of the smaller
ones (20 and 22). In the sequel we frequently use as working data set
a long contig in Chromosome 20 of the Homo Sapiens genome. 
This genomic contig is the locus N1\textunderscore 011387 (primary assembly) and 
contains 26259569 base pairs (bps), 
while the entire Chromosome 20 contains $\sim 63 \times 10^6$  bps. 
This contig is a DNA entity  long enough to ensure good statistics, when 
addressing both  the short and long range spatial properties. Moreover it 
is representative of the entire DNA molecule, since it contains both coding 
and non-coding sequences and other functional elements
 in similar densities as for all other human chromosomes.
In particular, the nucleotide frequencies for the contig are: 
$p_A=0.289341, p_C=0.208691, p_G=0.209448$ and $p_T=0.292519$.
 Occasionally, unknown
bps  denoted by $N$  which still resist in todays sequencing techniques,
 are found in genomes.  The $N$-percentage is very small and does not
contribute significantly to the statistics. We can then choose either to eliminate all $N$'s or to replace
them randomly with one of the other four $\{ A,C,G,T\}$. For both choices the presented results
are indistinguishable, up to insignificant statistical errors.
Very similar nucleotide frequencies are found in the other human chromosomes. The frequency of 
the four base pairs is not constant throughout the genome but varies along the 
sequences depending on evolutionary factors and on the presence (or absence) 
of functional units. For example the presence of the complex $CpG$ is associated 
with the presence of isochores, DNA regions with  high density of gene-coding regions. 
By coarse graining the alphabet at the PU-PY level,  the latter frequencies become very close: 
$p_{PU}=0.498788$ and $p_{PY}=0.501210$. Thus, information on possible presence of 
isochores faints. Alternatively, by refining the alphabet, for example by considering 
explicitly the frequencies of doublets, triplets etc.,  information on finer and finer 
scales emerge, which cannot be adherent from superpositions of previous levels of observation. 
This is one of the main elements
 which leads one to characterize this molecules as complex, since different levels 
of complexity appear when varying the scale of observation. 
\par For conciseness, we denote from now on purines and pyrimidines as states 1 and 2 respectively in
the two-letter alphabet and $A,C,G,T$ as states 1,2,3,4 respectively in the four-letter alphabet.
In Table \ref{table:2.1} the two-letter and four-letter conditional probabilities 
\begin{equation}
w_{ij}=W(i|j)=\frac{Prob(j,i)}{Prob(j)}
\label{eq01} 
\end{equation}
obtained by counting the frequencies of adjacent bps $i$ and $j$ averaged over the
entire sequence are provided. 
 Whereas $W$ in the two-letter case is nearly symmetric,
it is markedly asymmetric in the four-letter case. In particular,
$w_{32}$, the probability of encountering $G$ after $C$ is
noticeably smaller than the others. This difference
is well known in the biology literature and is attributed to the 
specific regulatory function of the $CpG$-complex 
in the human genome, being an
essential structural element of the promoters. In spite of such differences
all $w_{ij}$'s keep statistically significant values, suggesting that the
process underlying the entire structure is ergodic. Higher order probabilities
are obtained in a similar way (available upon request).

\begin{table}
\begin{center}
\begin{tabular}{|l|l|l|l|l|}
\hline
Two-letter alphabet & $w_{11}=0.559964$ & $w_{12}=0.440037$
& $w_{21}=0.437910$ & $w_{22}=0.562090$
\\ \hline\hline
\multirow{4}{3.5cm}{Four-letter alphabet} 
&$w_{11}= 0.324143 $& $w_{12}=0.173548 $&
$w_{13}=0.245801      $ & $w_{14}=0.256508$ \\ \cline{2-5}
&$w_{21}=0.3533467 $&$w_{22}= 0.259233 $&
$w_{23}=4.575338 E-002$ &$w_{24}= 0.341667$\\ \cline{2-5}
&$w_{31}=0.2869938 $& $w_{32}=0.210515 $&
$w_{33}=0.259181      $ &$w_{34}= 0.243310 $ \\ \cline{2-5}
&$w_{41}=0.2109355 $&$w_{42}= 0.206090 $&
$w_{43}=0.254663      $ & $w_{44}=0.328311$ \\ \hline
\end{tabular}
\caption{\label{table:2.1} Conditional probabilities $w_{ij}$ as obtained
from the DNA data.}
\end{center}
\end{table}

\section{Markov chain analysis}
\label{sec:markov}

\par
In the preceding section we have drawn inferences about probabilities
of various orders from a long, unbroken data set. These data are viewed as 
defining the states of an underlying system at points $n$ along the sequence, the succession
of which is supposed to be governed by a set of probability laws. The first question 
that comes then to the mind is whether these laws can be expressed for all practical purposes
in the form of a low order Markov chain.

\par A stochastic process $\{ i_n \} $  is a Markov chain of order $s$ if the conditional
probability 
\begin{equation}
W\left( i_n|i_1,i_2,\cdots i_{n-1}\right)=\frac {Prob\left( i_1,i_2,\cdots i_n \right) }
{Prob\left( i_1,i_2,\cdots i_{n-1}\right) }
\label{eq02}
\end{equation}
\noindent
is independent of $i_m$ for $m< n-s$. As stated in Sec. \ref{sec:data} it is understood throughout
that all these probabilities are considered as stationary, in the sense of being
independent of $n$. The simplest setting is that of a first order Markov chain, $s=1$.
Suppose that the conditional probability matrix $W(i|j)=w_{ij}$ has been evaluated
from some model. Estimating the singlet and doublet frequencies $np_i$ and
$np_{ij}$ from the data by different independent counts, leads then one 
to test the legitimacy of the model as a first order Markov chain on the basis
of the smallness of the differences $p_{ij}-
p_jw_{ij}$. A fundamental result in this context is that the random vector
\begin{equation}
(np_{ij}-np_jw_{ij})/(np_j)^{1/2}
\nonumber 
\end{equation}
converges to the normal distribution with covariance matrix determined by $w_{ij}$.
Keeping in mind the independence of the different samples, it follows \cite{billingsley:1961}
 that the sum
\begin{equation}
\sum_{ij}\frac{(np_{ij}-np_jw_{ij})^2}{(np_jw_{ij})}
\label{eq03}
\end{equation}
obeys asymptotically to the chi-square distribution. This opens the way to testing the
order of the Markov chain within a certain confidence interval by chi-square type tests.
These results can be extended rather straightforward to higher order Markov chains.
\par In many cases - including the problem addressed in this work -
one disposes of no reliable model for estimating, a priori, the conditional
 probabilities $w_{ij}$. The question thus arises whether
chi-square type tests for the order of the Markov chain can still be conducted on the
sole basis of the data. As suggested in refs. 
\cite{hoel:1954,lowry:1968,avery:1999}
 the answer is in the affirmative provided that the following chi-square
tests are used for the hypothesis that the chain is of the order $r$ 
\begin{equation}
\chi ^2=\sum_{i_1,i_2\cdots i_s}\frac{p_{i_1,i_2\cdots i_s}-p_{i_1,i_2\cdots i_{s-1}}
W(i_s|i_{s-r}\cdots i_{s-1})}{ p_{i_1,i_2\cdots i_{s-1}}
W(i_s|i_{s-r}\cdots i_{s-1})}          
\label{eq04} 
\end{equation}
\noindent where the $W$'s are estimated from the data as ratios of frequencies of
$i_1,i_2\cdots i_{s-1}$ and $i_1,i_2\cdots i_s$, $s$ being the maximum order considered.
\par To apply this test to our data we need to prescribe a confidence interval, 
which we have chosen to be 5\%, and compare the corresponding $\chi ^2$ value as given in the
Tables to the value \eqref{eq04}
obtained from the data. This requires specifying each time
the number of degrees of freedom, which is related to $s,r$ and the number of states $N$ by
\begin{equation}
\text{DOF = Number of degrees of freedom =}N^s-N^{s-1}-(N^r-N^{r-1})       
\label{eq05}
\end{equation}
The order of the process is then estimated to be the smallest
value of $r$ which produces a nonsignificant test statistics \cite{papapetrou:2013,menendez:2011}.

\begin{table}
\begin{center}
\begin{tabular}{|c|c|c|c|}
\hline
Orders compared& DOF& $\chi ^2$-value \eqref{eq04}
& $\chi ^2$-value at 5\% level \\ \hline
0 1 & 1 & $0.391189\times 10^{6}$ & 3.84 \\\hline
1 2 & 2 & $0.936032\times 10^{5}$ & 5.99\\ \hline
2 3 & 4& $0.840413\times 10^{4}$  & 7.81\\\hline
3 4 & 8 & $0.244684\times 10^{5}$  & 9.48\\\hline
4 5 & 16 & $0.341158\times 10^{5}$ & 11.07 \\\hline
\end{tabular}
\caption{\label{table:3.1} $\chi ^2$-test \eqref{eq04} for the DNA data
in the two-letter alphabet. }
\end{center}
\end{table}

\begin{table}
\begin{center}
\begin{tabular}{|c|c|c|c|}
\hline
Orders compared& DOF & $\chi ^2$-value (3.3) & $\chi ^2$-value at 5\% level \\ \hline
0 1 & 9 & $0.143217\times 10^{7}$ & 17 \\\hline
1 2 & 36 & $0.361137\times 10^{6}$ & 51\\ \hline
2 3 & 144& $0.165965\times 10^{6}$  & 150\\\hline
3 4 & 576 & $0.227638\times 10^{6}$  & 633\\\hline
4 5 & 2304 & $0.322366\times 10^{6}$ & 2417 \\\hline
\end{tabular}
\caption{\label{table:3.2} $\chi ^2$-test \eqref{eq04} for the DNA data in
the four-letter alphabet.}
\end{center}
\end{table}

Tables \ref{table:3.1} and 
\ref{table:3.2} summarise the results from our data obtained using the chi-square test
for the two- and the four-letter alphabets, respectively. In all cases tested the $\chi ^2$ values
obtained by applying \eqref{eq04}
to the data turn out to be
 much larger than the confidence level ones for processes
of order up to 6. In other words, the DNA data cannot be fitted by a low order Markov chain. 
For comparison, by simulating a first order Markov chain having the same $p_i$'s and $w_{ij}$'s
as the data and by applying the test leads to a value of $0.182155\times 10^6$ for the first 
row of Table \ref{table:3.1} and subsequently for the second raw to a value $0.392362\times 10^1$, smaller
than the $\chi ^2$ value of 5.99 at 5\% level.

\section{Spatial asymmetry and probability fluxes}
\label{sec:fluxes} 

\par The failure of the Markov property established in the preceding section 
leads us to search for alternative ways to characterize DNA viewed as a symbolic 
sequence, or alternatively, as a text written in the four-letter alphabet 
provided by the four nucleotides or in the restricted two-letter alphabet of 
purines and pyrimidines. A common syndrome of all languages is irreversibility 
in the form of “spatial” asymmetry, i.e., reading a text written in the language 
from, say, left to right produces a different result from reading it from right to left.
In DNA sequences such spatial asymmetries may be expected due to the 
extensive presence of repetitive elements and to large scale patchiness \cite{polak:2009,oliver:2008}.
 In this section we address the issue of 
irreversibility and asymmetry for the DNA on the basis of data summarized in Sec. \ref{sec:data}.

	At the microscopic level of description irreversibility and asymmetry are 
associated with the breakdown of the property of detailed balance, i.e., that in a 
given system the probability of an event leading from an initial state $i$ to a final 
state $j$ is counteracted by the probability of the reverse event leading from state
 $j$ to state $i$. This implies, in turn, the presence of a global constraint keeping the 
system out of the state of thermodynamic equilibrium. Transposed 
 from the time domain to the one of the DNA symbolic sequence as it unfolds in space the 
simplest expression of this property amounts to the joint probability of two states $i$
 and $j$ in adjacent positions $n$ and $n+1$ along the chain satisfying the “space reversal relation”:
\begin{subequations}
\begin{align}
p(i,n;j,n+1) =p(j,n;i,n+1)\hspace{5cm}
\label{eq06a} \\
\text{or using the definition of conditional probabilities, }\hspace{6cm}
\nonumber \\
w_{ji}p_i  =w_{ij}p_j \hspace{6cm}
\label{eq06b}
\end{align}
\end{subequations}

\noindent
Here $i,j$ run from 1 to 4 in the case of the four-letter alphabet defined by the nucleotides 
and from 1 to 2 for the two-letter purine-pyrimidine alphabet. Alternatively, the probability flux $J^{(2)}_{ij}$
\begin{equation}
J^{(2)}_{ij}=w_{ji}p_i-w_{ij}p_j
\label{eq07}
\end{equation}
vanishes if the detailed balance condition is satisfied. In a similar vein, higher order space reversal
 conditions involving more than two sites can be introduced, e.g.
\begin{subequations}
\begin{align}
P(i,n;j,n+1;k,n+2)=p(k,n;j,n+1;i,n+2)\hspace{4cm}
\label{eq8a}\\
\text{or equivalently,}\hspace{14cm}\nonumber \\
w_{kji}p_{ij}-w_{ijk}p_{kj}=p_iw_{ji}w_{kji}-p_kw_{jk}w_{ijk}=0\hspace{5cm}
\label{eq08b}
\end{align}
\end{subequations}
\noindent
expressing the vanishing of the probability flux
\begin{equation}
J^{(3)}_{ijk}=w_{kji}p_{ij}-w_{ijk}p_{kj}
\label{eq09}
\end{equation}
Notice that for an alphabet of more than two letters there are more than one probability
fluxes and more than one detailed balance conditions. For instance in the four-letter 
alphabet there are six fluxes $J^{(2)}_{ij}$. If the process were first order Markov,
these fluxes would be related by the stationarity condition:
\begin{equation}
p_{j}=\sum_{i}w_{ji}p_i
\nonumber
\end{equation}
or, using the normalisation property $\sum_{i}w_{ij}=1$ ,
\begin{equation}
\sum_i \left( w_{ji}p_i-w_{ij}p_j \right) =\sum_{i}J^{(2)}_{i j}=0
\label{eq10}
\end{equation}
 There would then be only three independent fluxes, say $J^{(2)}_{12}, J^{(2)}_{13}$ and 
$J^{(2)}_{23}$. One may also define composite fluxes, e.g. the flux from state 1 to the
pyrimidines, 
\begin{equation}
J_{1,PY }=\left( w_{21}p_1-w_{12}p_2 \right)+ 
\left( w_{41}p_1-w_{14}p_4 \right) =J^{(2)}_{12}+J^{(2)}_{14}
\label{eq11}
\end{equation}
 and the purine-pyrimidine flux as

\begin{align}
J_{PU,PY }&=J^{(2)}_{12 }+J^{(2)}_{14 }+J^{(2)}_{32 }+J^{(2)}_{34 } \nonumber \\ 
             &= J_{1,PY }+J_{3,PY }
\label{eq12}
\end{align}
\noindent
The latter would be strictly zero had the process been a first order Markov.

 We now
proceed to the evaluation of the probability fluxes from the data and to the 
testing of the detailed balance condition.  Table \ref{table:4.1} summarises the main result for fluxes
 $J^{(2)}$ and $J^{(3)}$ in the case of the two-letter alphabet. As can be seen, the fluxes are very
small. Actually they are indistinguishable from those  obtained from a random sequence of the same length
and with probabilities $p_i$ fitted from the data
(not shown). Detailed balance holds therefore true in this case,
or, to put it differently, there is no overall spatial asymmetry and irreversibility. This is compatible
with the symmetry of the associated conditional probability matrix pointed out in Sec. \ref{sec:data}.

\begin{table}
\begin{center}
\begin{tabular}{|c|c|}
\hline
$J^{(2)}$ &  \\ \hline
12 & $2.9802\times 10^{-8}$  \\\hline\hline
$J^{(3)}$ &  \\ \hline
112 & $5.2154\times 10^{-8}$  \\\hline
212 & $3.7252\times 10^{-8}$  \\\hline
\end{tabular}
\caption{\label{table:4.1} Probability fluxes $J^{(2)}$ and $J^{(3)}$ in the two-letter alphabet}
\end{center}
\end{table}

Table \ref{table:4.2}
 summarises the results for $J^{(2)}$ and $J^{(3)}$ in the case of the four-letter alphabet.
The results  are now definitely significant, much larger than those obtained
from a random sequence. We conclude that detailed balance does not hold here, in other words, 
there is an overall irreversibility in the form of spatial asymmetry.
 Following the analogy with the theory
of stochastic processes and nonequilibrium statistical mechanics, one would be tempted to state that
the structure as a whole can be viewed as a system subjected to constraints maintaining it far from
the state of thermodynamic equilibrium. This may, at a first sight, sound in contradiction with the
usual view of DNA as a stable molecule in thermodynamic 
equilibrium with its environment. But the contradiction is only
apparent inasmuch as complex matter in general, and DNA in particular as we observe it today, is 
to be viewed as the outcome of a  primordial nonequilibrium evolutionary process that was
eventually stabilised in a ``fossil'' form as a result of the action of local short-ranged intermolecular
interactions. Otherwise, the waiting time to see this event happen spontaneously would be
exceedingly large
owing to the combined effects of detailed balance and of the explosion of the number of
possible combinations of the constituting subunits among which only a small subset would
posses the desired biological functions \cite{nicolis:1989,nicolis:2012,frisch:1984}.

\begin{table}
\begin{center}
\begin{tabular}[t]{|c|c|}
\hline
$J^{(2)}$ &  \\ \hline
12 & $-2.3525\times 10^{-2}$  \\\hline
14 & $1.2515\times 10^{-2}$  \\\hline
32 & $3.4543\times 10^{-2}$  \\\hline
34 & $-2.3533\times 10^{-2}$  \\\hline
\end{tabular}
\hspace{2cm}
\begin{tabular}{|c|c|}
\hline
$J^{(3)}$ &  \\ \hline
123 & $1.1347\times 10^{-2}$  \\\hline
124 & $7.7443\times 10^{-3}$  \\\hline
134 & $6.6033\times 10^{-4}$  \\\hline
213 & $-1.2467\times 10^{-2}$  \\\hline
214 & $-3.7790\times 10^{-3}$  \\\hline
234 & $1.2393\times 10^{-2}$  \\\hline
312 & $1.2358\times 10^{-2}$  \\\hline
314 & $-3.6969\times 10^{-3}$  \\\hline
412 & $7.8714\times 10^{-4}$  \\\hline
413 & $7.5745\times 10^{-3}$  \\\hline
\end{tabular}
\caption{\label{table:4.2} Probability fluxes $J^{(2)}$ and $J^{(3)}$ in the four-letter alphabet}
\end{center}
\end{table}

\section{Entropy analysis and information transfer}
\label{sec:entropy}

\par In this section we introduce and analyse a  set of quantities aiming to characterise
the complexity of the DNA symbolic sequence, while accounting for its central role as information
source 
\cite{ebeling:1991,nicolisjs:1991,roman:1996} as well as for 
the spatial asymmetry and irreversibility established in the preceding 
section. The simplest quantity in this family is the information (Shannon) entropy:
\begin{equation}
S_I=-\sum_ip_i\ln p_i 
\label{eq13}
\end{equation}
\noindent describing the amount of choice exercised by the information source and the associated
uncertainly of the message recipient. By its static character this quantity does not provide
insights on the overall structure of the sequence. To handle this aspect we divide the sequence into
blocks of symbols $i_1,i_2\cdots i_n$ of length $n$ and extend  
Eq. \eqref{eq13}, which defines essentially
the entropy per symbol, to the entropy per block of symbols over a window of length $n$:

\begin{equation}
S_n=-\sum_{i_1,i_2,\cdots i_n}P(i_1,i_2,\cdots i_n)\ln P(i_1,i_2,\cdots i_n)       
\label{eq14}
\end{equation}
where the sum runs over all sequences compatible with the underlying rules.

\par Now suppose that the source has sent a message in the form of a particular $n$-sequence. 
What is the probability that the next symbol be $i_{n+1}$? Clearly we are dealing here with a 
conditional event. The entropy excess associated with the addition of an extra symbol to the 
right of the $n-$block (``word'') is then \cite{nicolis:2012,ebeling:1991}:

\begin{equation}
h_n=-\sum_{i_1,i_2,\cdots i_n,i_{n+1}}P(i_1,i_2,\cdots i_n)W(i_{n+1}|i_1,i_2,\cdots i_n)
\ln W(i_{n+1}|i_1,i_2,\cdots i_n)       
\label{eq15}
\end{equation}

The first nontrivial value $h$ of $h_n$ describes the amount of information obtained when
one moves along the chain one step ahead of the initial state $i_1$,

\begin{equation}
h=h_1=-\sum_{i,j}p_iw_{ji}\ln w_{ji}      
\label{eq16}
\end{equation}
Actually, $h_n$ would be $n-$independent and equal to $h$ had the sequence been compatible 
with the Markov property, which, as shown in Sec. \ref{sec:markov}, is not the case. 
In spite of this failure, \eqref{eq16} keeps its
significance whatever the nature of the process and will be referred to, in the sequel, as the
Kolmogorov-Sinai (KS) entropy. To capture the asymmetry property analysed in Sec. \ref{sec:fluxes}
 it is also
useful to introduce the reverse process in which the order of the states visited is
running backwards, and to define the associated KS entropy as

\begin{equation}
h^R=-\sum_{i,j}p_i w_{ji}\ln w_{ij}       
\label{eq17}
\end{equation}
One can easily check that if the Markov property holds $h^R$ is larger than or equal to $h$. Indeed,

\begin{equation}
\sigma_I=h^R-h=\sum_{ij}p_iw_{ji}\ln \frac{w_{ji}}{w_{ij}}       
\label{eq18}
\end{equation}
or using the normalisation and stationarity properties discussed in Sec. \ref{sec:fluxes} 
\cite{nicolis:2012,gaspard:2004,luo:1984,andrieux:2008},

\begin{equation}
\sigma_I=h^R-h=\frac{1}{2}\sum_{ij}\left( w_{ji}p_i-w_{ij}p_j\right) 
\ln \frac{w_{ji}p_i}{w_{ij}p_j} \ge 0      
\label{eq19}
\end{equation}

\par We can express this property by the statement that the direct sequence is more ordered than the 
reverse one as long as the probability flux $J^{(2)}_{ij}$ (Eq. \eqref{eq07})
does not vanish,
 i.e., as long as detailed balance does not hold. For this reason we will refer to $\sigma _I$, which can be regarded as a
distance from the regime of detailed balance, as the {\it information entropy production}. 
Conversely, in absence of the Markov property but knowing that $J^{(2)}_{ij}$ is different from zero, one
may wonder whether $h^R$ is still larger than $h$. As we see shortly, this is indeed the case for the
DNA sequence in the four-letter alphabet.
\par
\begin{figure}

\includegraphics[clip,width=0.45\textwidth,angle=0]{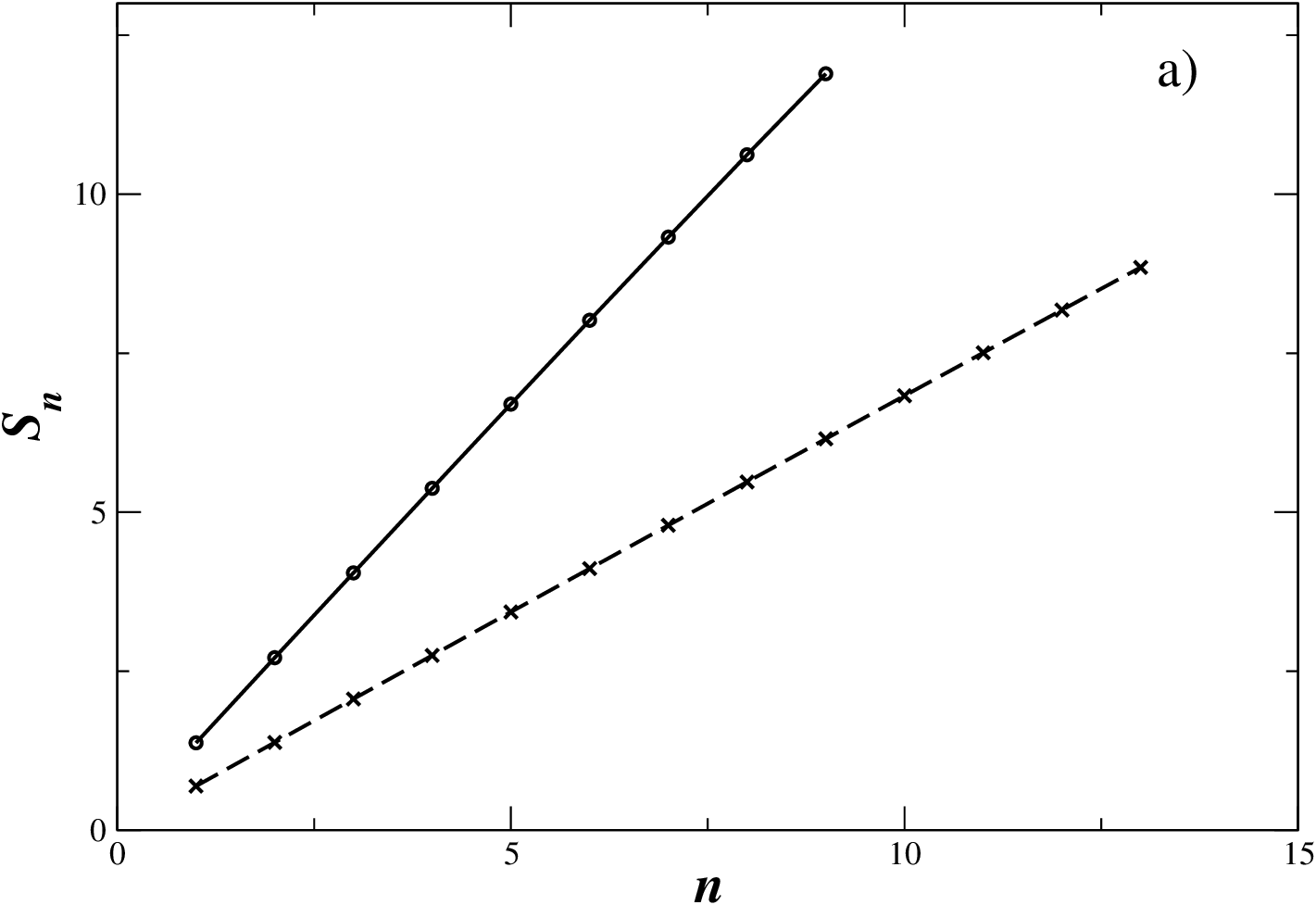}
\includegraphics[clip,width=0.49\textwidth,angle=0]{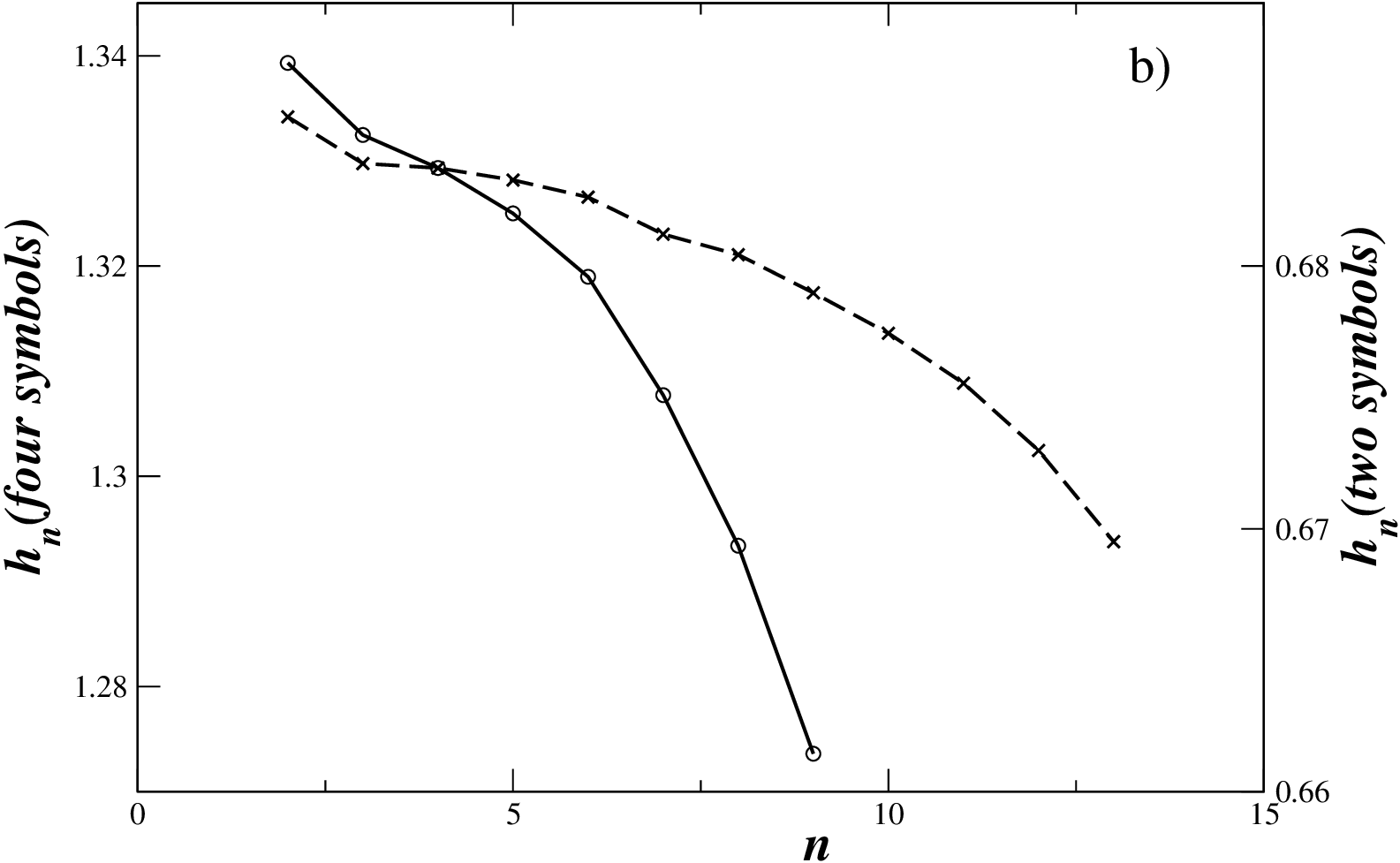}

\caption{\label{fig:01}{}  a) Information entropy $S_n$ 
for blocks of size $n$ for 
the DNA data in the four- and two-letter alphabets; b) Entropy excess $h_n$ for block of size $n$ for 
the DNA data in the four- and the two-letter alphabets.\hspace{4cm}} 
\end{figure}

 Let now $n$ be gradually increased. As stated earlier in a Markov process $h$ would 
remain constant, entailing that $S_n$ would increase linearly in $n$. 
Figure \ref{fig:01} depicts the dependence of $S_n$
and $h_n$ as defined from Eqs. \eqref{eq14}-\eqref{eq15}
for the DNA data of Sec. \ref{sec:data}. 
 As can be seen the $S_n$ versus $n$ dependence is not strictly linear. $h_n$ varies thus
(weakly but systematically) with $n$ from a value 1.339 for $n=1$ to 1.273 for $n=8$,
in the case of the four-letter
alphabet (solid lines in Fig. \ref{fig:01}). This is in
accord with the conclusion drawn in Sec. \ref{sec:fluxes}
on the non -Markovian character of the sequence and suggests the presence of long-range correlations
(see also Sec. \ref{sec:exit-distances} below).
\par As expected the value $h_1$ reported above is identical to the value $h$ as computed 
directly from the data. Table \ref{table:5.1} summarises the results of evaluation of $h$, $h^R$ and $\sigma_I$
using Eqs. \eqref{eq16}-\eqref{eq18},
for the four-letter
alphabet. For comparison the corresponding values from a random sequence of the
same length are also given. 
In this case, as expected $h$ and $h^R$ are both equal to the maximum entropy, $h=h^R=\ln 4$, of the
sequence and $\sigma_I=0$. 

\begin{table}
\begin{center}
\begin{tabular}{|c|c|c|c|}
\hline
 & $h$ & $h^R$ & $\sigma_I$\\ \hline
DNA Data & 1.339 & 1.416 & 0.077 \\ \hline
Random sequence& 1.373 & 1.373& 4$\times 10^{-7}$ \\\hline
\end{tabular}
\caption{\label{table:5.1} Kolmogorov entropy of the direct and
reverse sequence and information entropy production as
computed from the DNA data. \hspace{4cm} }
\end{center}
\end{table}

\par The evaluation of the quantities in Table \ref{table:5.1} 
for the DNA sequence in the two-letter alphabet
leads to the quite different conclusion that $h\sim h^R=0.686$ and thus $\sigma_I\approx 0$, the
corresponding $h-$value for the random sequence being $h\sim \ln 2=0.693$. 
On the other hand $S_n$ and $h_n$ still depend on $n$ in a non-trivial way, see 
dashed lines in Fig. \ref{fig:01}.
The signature of the
non-Markovian character at the level of the entropy
quantities therefore persists in the two-letter alphabet case.
\par
On the basis of the above comparison between the two alphabets and between the DNA data and those
associated to the random sequence, one is tempted to conclude that revealing the asymmetry of the
DNA sequence in the four-letter alphabet - as established already in Sec. \ref{sec:fluxes}
 - has also some
interesting signatures at the level of information processing: Information is being produced
(in the sense $\sigma_I >0$)
as long as one advances along a preferred direction in sequence space, and this requires reading
the ``text'' in a four-letter alphabet.

\begin{figure}
\includegraphics[clip,width=0.6\textwidth,angle=0]{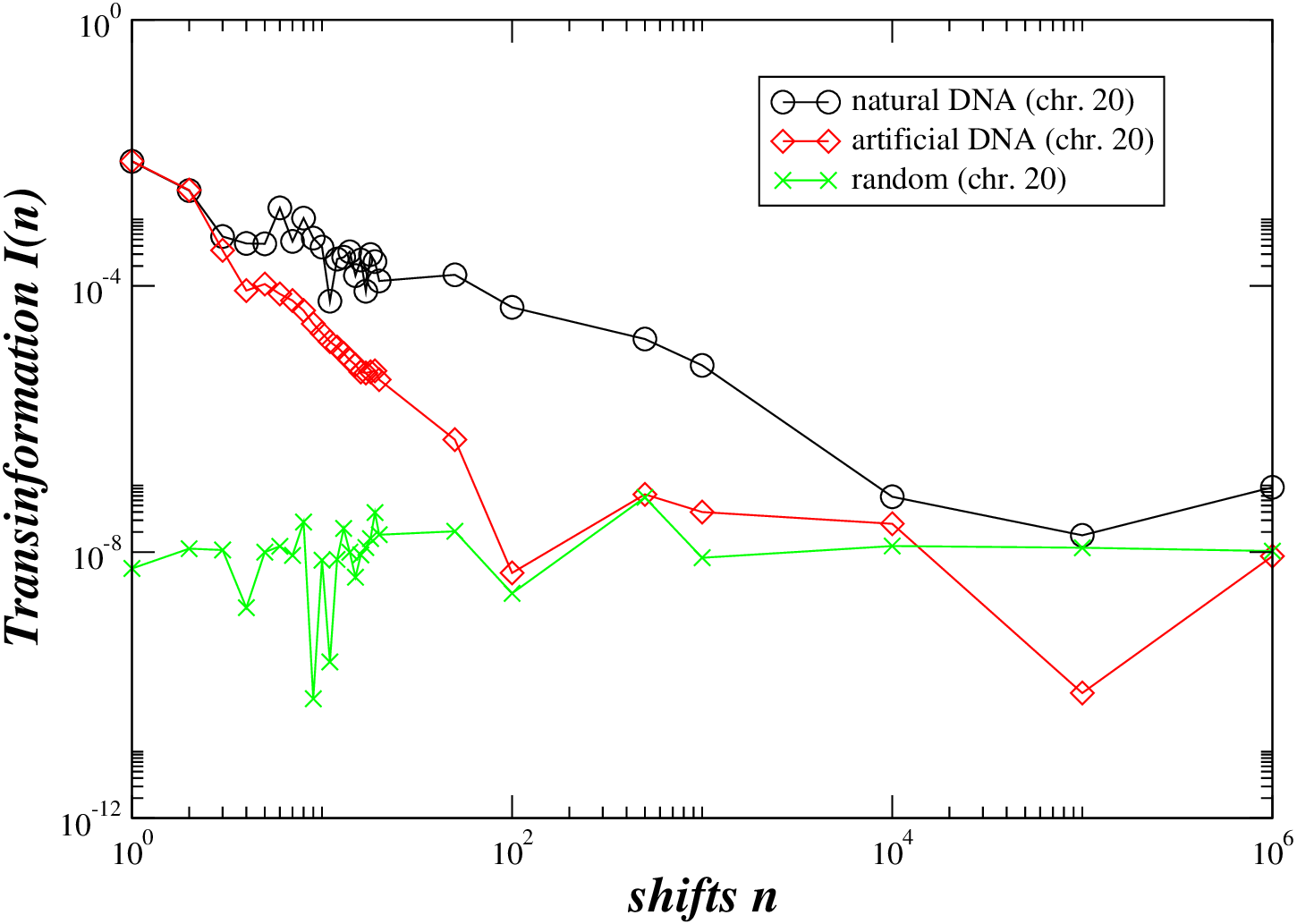} 

\caption{\label{fig:02}{} (Colour online) Information transfer $I(n)$ between a
sequence and its shift by $n$ symbols versus $n$. Line with circles depicts the DNA sequence,
line with crosses a random sequence and line with diamonds a model-generated sequence 
(Sec. \ref{sec:exit-distances}).
}
\end{figure}

An alternative view of the DNA sequence in connection with both
the presence of correlations and  information processing is to 
consider two segments  - typically of the same length - view the leftmost segment
(say $x$) as the ``source'' and the
second one (say $y$), as the ``receiver'' and evaluate the information transfer from $x$ to $y$.
We define this quantity as \cite{nicolisjs:1986}:
\begin{equation}
I_{x\to y}=\sum_{i,j}P(x_i,y_j)
\ln \frac{P(x_i,y_j)}{P(x_i)P(y_j)}.       
\label{eq20}
\end{equation}
We notice that
\begin{equation}
I_{x\to y}=S_I(y)+\sum_{i,j}P(x_i,y_j)\ln W(y_j|x_i).      
\nonumber
\end{equation}
Furthermore, if $x$ and $y$ are
 two adjacent sites of the sequence the second term, which represents the
conditional entropy of $y$ given the state of $x$, reduces to the KS entropy (Eq. \eqref{eq16}).
In the following analysis a sequence is compared with its shifts. For a specific shift of $n$
sites, the information
transfer represents then the capacity between adjacent symbols
to interact down the sequence.  The upper
line in Fig. \ref{fig:02}
depicts the dependence of $I_{x\to y}$ on $n$
for two sequences of the same length: the chromosome 20, 
working contig N1\textunderscore 011387 (in two-letter representation) and its shift by
1,2, $\cdots$ up to $10^5$ symbols. The last excess $n$ symbols in the comparison can either be
reinjected at the beginning of the sequence, or they can be neglected without
changing significantly the resulting $I$ values.
 For comparison, the lower line (with crosses) stands for the 
results obtained from random sequences of the same length and bps frequencies. The intermediate
line (with diamonds) is associated with the model which will 
be discussed in Sec. \ref{sec:model}. As can be
seen from the figure the information transfer as extracted from the DNA data remains higher with
respect to the case of a random sequence for shifts up to 100, suggesting,
once again, the presence of correlations and information transfer between
successive bps up to the order of $\sim 100$. At the level of functionality, this nontrivial
information transfer may refer to cooperations between successive units related to the presence
of codons in the coding regions and to the multiple presence of poly-A's, to frequent appearance
of repetitive elements, to the regulatory elements, to the promoters 
and to other functional elements
in the non-coding parts \cite{kalkatawi:2012,lustig:1984}.

A different view of the presence of correlations in information transfer between
two symbol sequences is provided
by their Hamming distance, which determines the number of positions at which the 
corresponding symbols are different or counts the number of substitutions required
 to change one sequence into the other. The classical
Hamming distance $H_{1-2}$ between the two symbol sequences $S1$ and $S2$, as 
defined by R. Hamming in 1950  for error detection, is \cite{hamming:1950}: 
\begin{equation}
 \begin{array}{l l}
H_{1-2}=\sum_{i}d_i, & \quad \text{where }\\ 
 d_i  = \left\{ 
  \begin{array}{l l}
    0 & \quad \text{if $S1(i)= S2(i)$}\\
    +1& \quad \text{if $S1(i) \not= S2(i)$}
  \end{array} \right.
   \end{array}
\end{equation}
We shall also use a modified Hamming distance $HM_{1-2}$ between $S1$ and $S2$,
defined as 
\begin{equation}
  \begin{array}{l l}
HM_{1-2}=\sum_{i}d_i, & \text{where} \\
 d_i = \left\{
  \begin{array}{l l}
    0 & \quad \text{if $S1(i)=S2(i)$}\\
    +0.5 & \quad \text{if $S1(i)$ and $S2(i)$ are both purines}\\
    +0.5 & \quad \text{if $S1(i)$ and $S2(i)$ are both pyrimidines}\\
    +1 & \quad \text{otherwise}
  \end{array} \right.
\end{array}
\end{equation}
\noindent which considers the purine-purine variation as less important than
the purine-pyrimidine one and penalises by 0.5 the $PU-PU$ discrepancy and 
by 1 the $PU-PY$ difference.
\par
\begin{figure}
\includegraphics[clip,width=0.45\textwidth,angle=0]{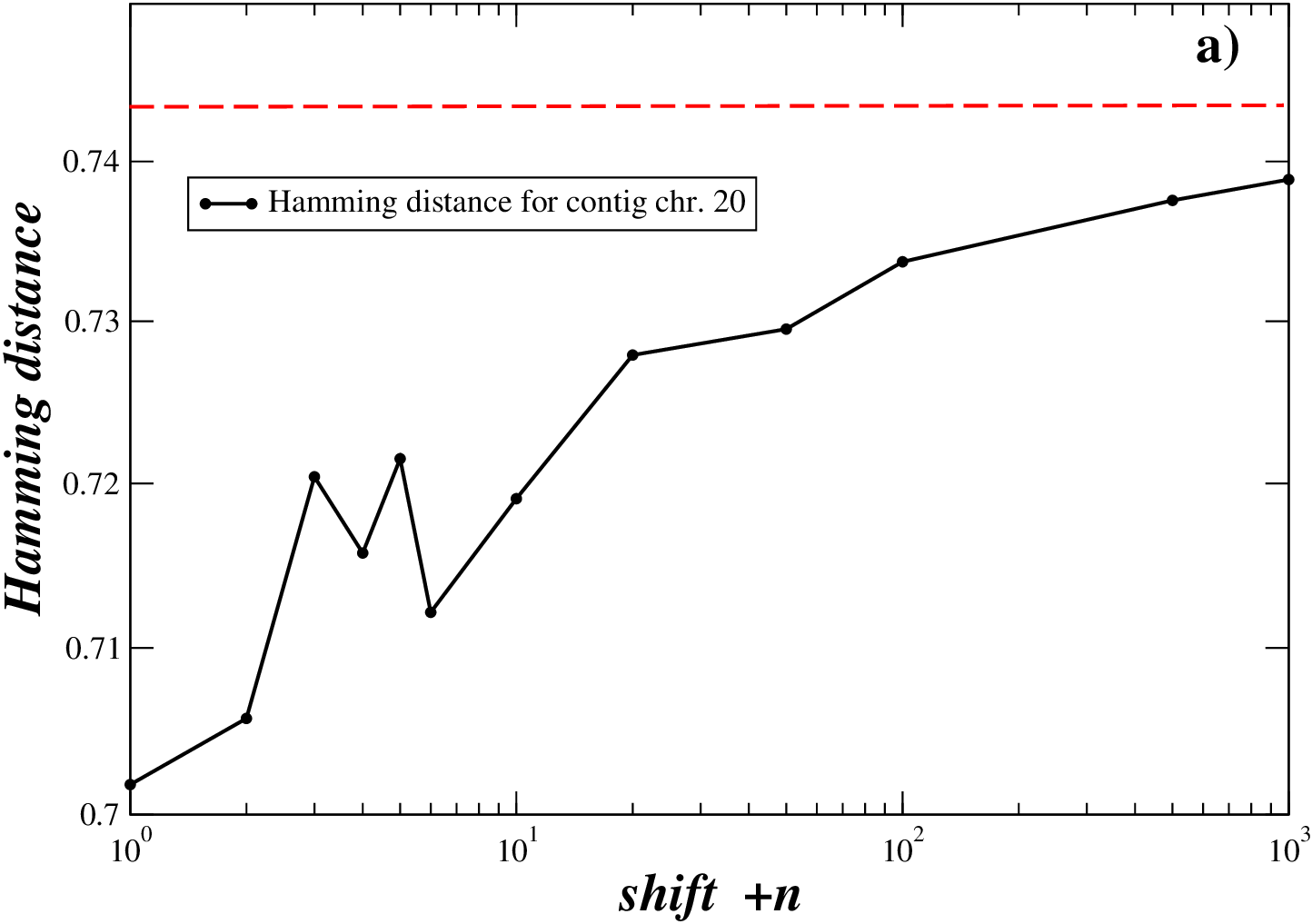}
\includegraphics[clip,width=0.45\textwidth,angle=0]{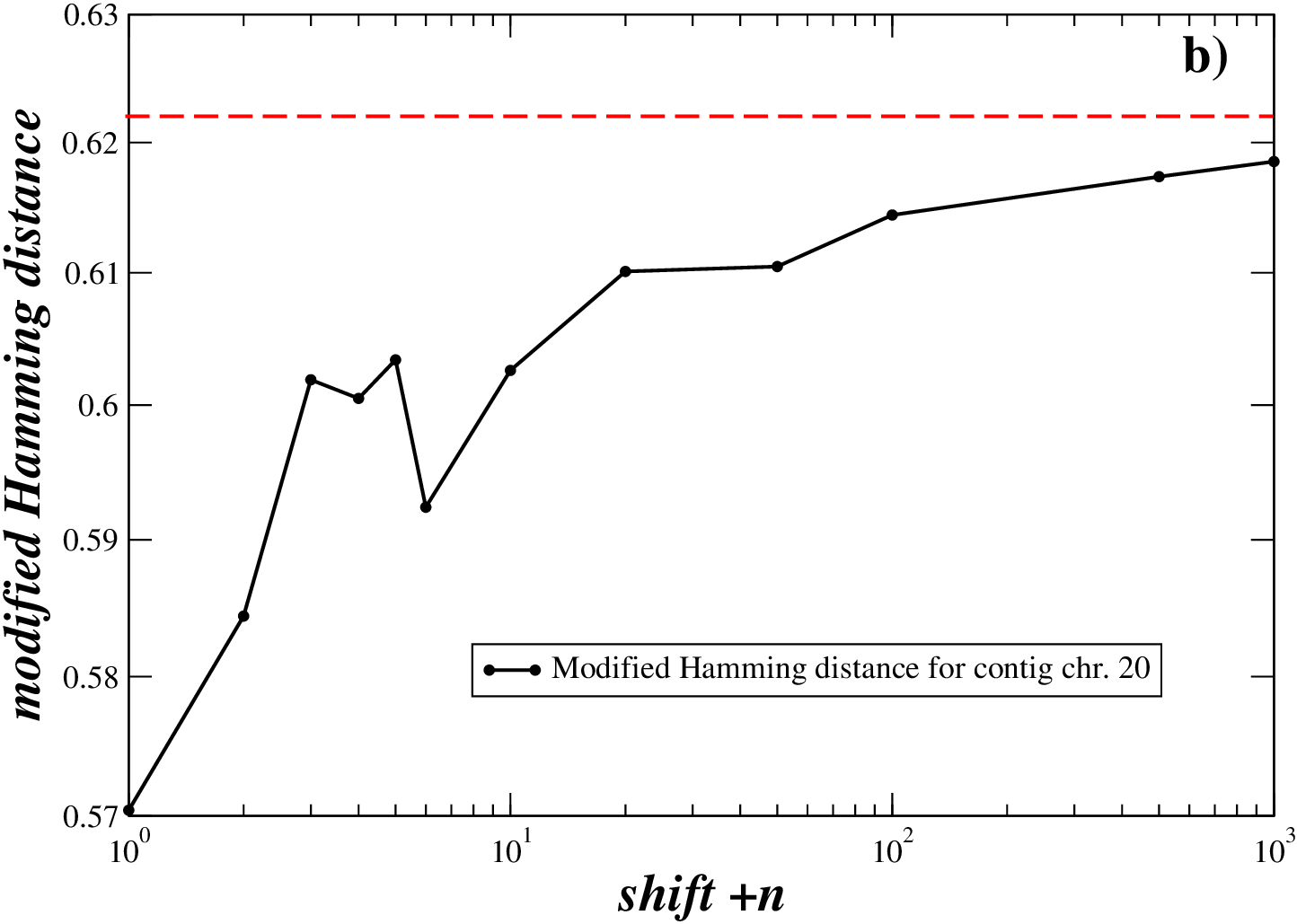}

\caption{\label{fig:03}{} (Colour online) a) The classical Hamming distance
$H_{1-2}$ between the contig sequence and its shifts. The red dashed line
denotes the
$H-$value between two random sequences with the same
symbol frequencies.
 b) The modified Hamming distance
$HM_{1-2}$ between the contig sequence and its shifts. The red dashed line
denotes the
$HM-$value between two random sequences with the same
symbol frequencies. All sequences are represented in the four-letter alphabet.
}
\end{figure}

In the case of the original Hamming distance, the $H-$distance between the contig
sequence and a random one with the same symbol frequencies is $H(contig-random)=0.74331$.
Note that if the 4 symbol frequencies were equal the value would be $12/16=0.75$.
The $H-$value between two random sequences are of the same order $H(random1-random2)=0.74329$.
The $H-$value between the contig sequence
and its shifts shows, on the contrary, interesting correlations similar to the ones demonstrated by the 
information transfer, see Fig. \ref{fig:03}a.

In the case of the modified Hamming distance, the $HM-$distance between the contig
sequence and a random one with the same symbol frequencies is $HM(contig-random)=0.6216$.
The $H-$value between two random sequences are of the same order $HM(random1-random2)=0.6220$.
 Again, the $HM-$value between the contig sequence
and its shifts shows interesting correlations, similar to the ones demonstrated by the 
information transfer and the original $H$-distance, see Fig. \ref{fig:03}b. 

\par Figures \ref{fig:03}a,b demonstrate overall
 that there is a non-trivial information flow between each
bps and its neighbours, while this information decreases as the bps become more and more
distant on the chain. 

\section{Exit and recurrence distance distributions}
\label{sec:exit-distances}

\par 
So far we have been concerned with global properties of the DNA sequences. In this section we
introduce a new set of quantities which allow probing features associated with 
the local structure. One example of special importance is the appearance of clusters in which a given
symbol - or a given subsequence of symbols - is repeated for a certain number of steps, beyond which a
transition to different symbols or subsequences is taking place.
\par To capture such features we introduce the exit distance
distribution, a concept analogous to the 
exit time distribution familiar from the theory of stochastic processes
\cite{feller:1968,gardiner:1983}. Specifically, starting with
a certain state/symbol $j$, we ask for the probability $q_{j,n}$ that an 
escape from it occurs at the n-th step as one moves along the sequence \cite{bala:1997}:

\begin{equation}
q_{j,n}=Prob(j,1:\overline j,n) /Prob(j,1)      
\label{eq21}
\end{equation}

\noindent where $\overline j$ denotes the set of all allowed states with the exception of $j$.

\par Closely related to the above is the recurrence distance distribution in which starting again with a
state $j$ we ask what is the probability $F_{j,n}$ to encounter this state again $n$ steps down the
sequence, with the understanding that the sites between 1 and $n$ are found in states $\overline j$,
other than $j$:
\begin{equation}
F_{j,n}=Prob(j,1;\overline j,2;\cdots ;j,n) /Prob(j,1)      
\label{eq22}
\end{equation}
For a first order Markov process both $q_{j,n}$ and $F_{j,n}$ can be evaluated explicitly on the sole
basis of the conditional probability matrix $W=\{ w_{ij}\} $. Specifically,
\begin{equation}
q_{j,n}=(w_{jj})^{n-1}-(w_{jj})^{n}     
\label{eq23}
\end{equation}
and $F_{j,n}$ is expressed in terms of its generating function $\tilde F_{j}(s)$ as
\begin{equation}
\tilde F_{j}(s)=\left[ sW(I-sW)\right]^{-1}     
\label{eq24}
\end{equation}
where $I$ is the unit matrix.  As a corollary, both $q_{j,n}$ and $F_{j,n}$ are superpositions of 
exponentials in $n$ and thus fall off exponentially for large $n$.

\par Equations \eqref{eq23}-\eqref{eq24}
can be extended rather straightforwardly to the case of a second-order 
Markov process. The calculations are more involved, but the property of exponential decay for 
large $n$ is again found to hold here.

\par
\begin{figure}
\includegraphics[clip,width=0.45\textwidth,angle=0]{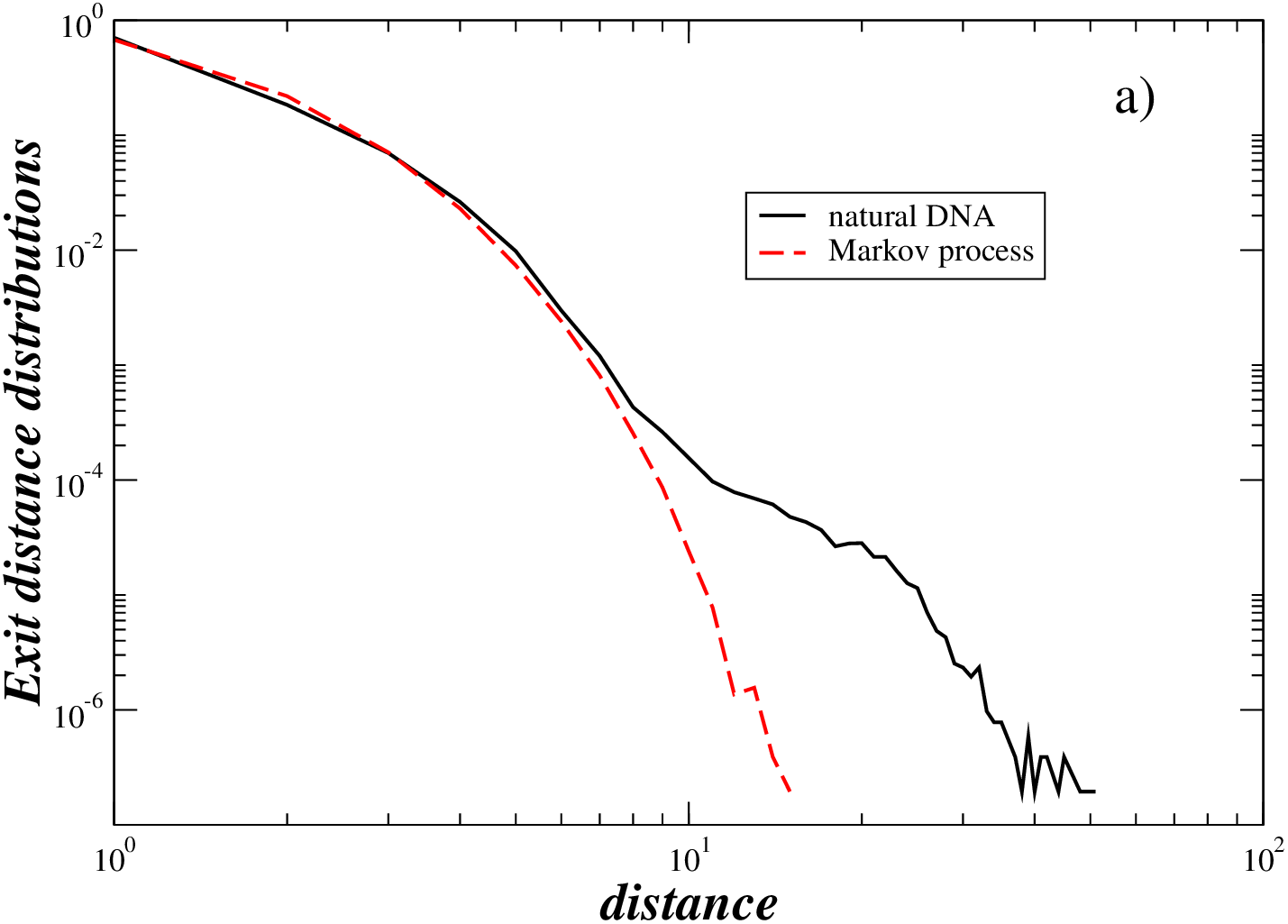}
 \includegraphics[clip,width=0.45\textwidth,angle=0]{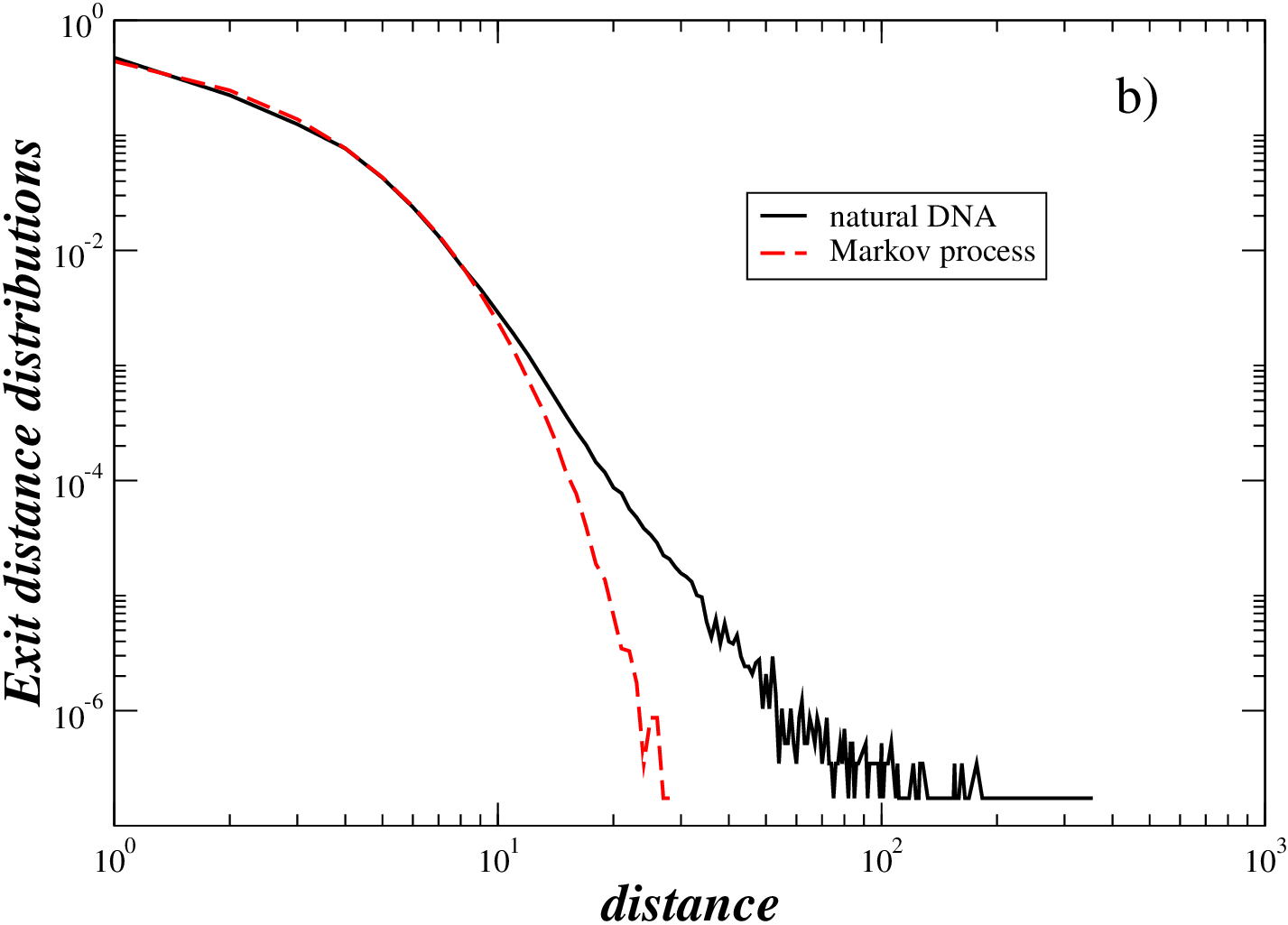} 
\caption{\label{fig:04}{} (Colour online) 
Exit distance distributions: a) four-letter alphabet, $C-$state; b) two-letter alphabet, 
$PU-$state.
\hspace{10cm}
 }
\end{figure}

 We now proceed to the evaluation of these distributions from the DNA contig data  in Sec. \ref{sec:data},
starting with $q_{j,n}$. For this purpose the data is being read along the direct sequence.
Whenever a state $i$ is first spotted the origin of coordinates is set on the corresponding site
and the distance from the origin is recorded when a state different from $i$ first appears along
the sequence. Counting all the distances recorded in this way for each of the states one arrives
at the exit distance distribution.   
In Figs. \ref{fig:04}a,b the distributions for state $C$ and for $PU$
 in the case of the four- and
the two-letter alphabets, respectively, are depicted. 
In both cases we observe the tendency for development of
 long tails (see also refs. \cite{masoliver:1986,altmann:2012}).
 In particular in the case of the four-letter alphabet the state $C$ shows a linear
region of low slope ($\sim -2$) in the intermediate scales
which is soon covered by finite size effects. The tendency for development of long tails is
better detected from comparison
with the associated probability for a first order Markov process indicated in the figures by the dashed
lines as obtained from a direct simulation of a Markov chain with conditional probabilities equal to
those provided by the data. Interestingly, the exit distributions from states  $A$ and $T$
 and from states
$C$ and $G$ are indistinguishable. Furthermore, the distribution of $A$ and $T$
 is longer-ranged
 than the one of $C$ and $G$
(not shown), owing principally to the existence of poly(A) and poly(T) 
domains found in the human genome \cite{kalkatawi:2012,lustig:1984}.
\par
\begin{figure}
\includegraphics[clip,width=0.6\textwidth,angle=0]{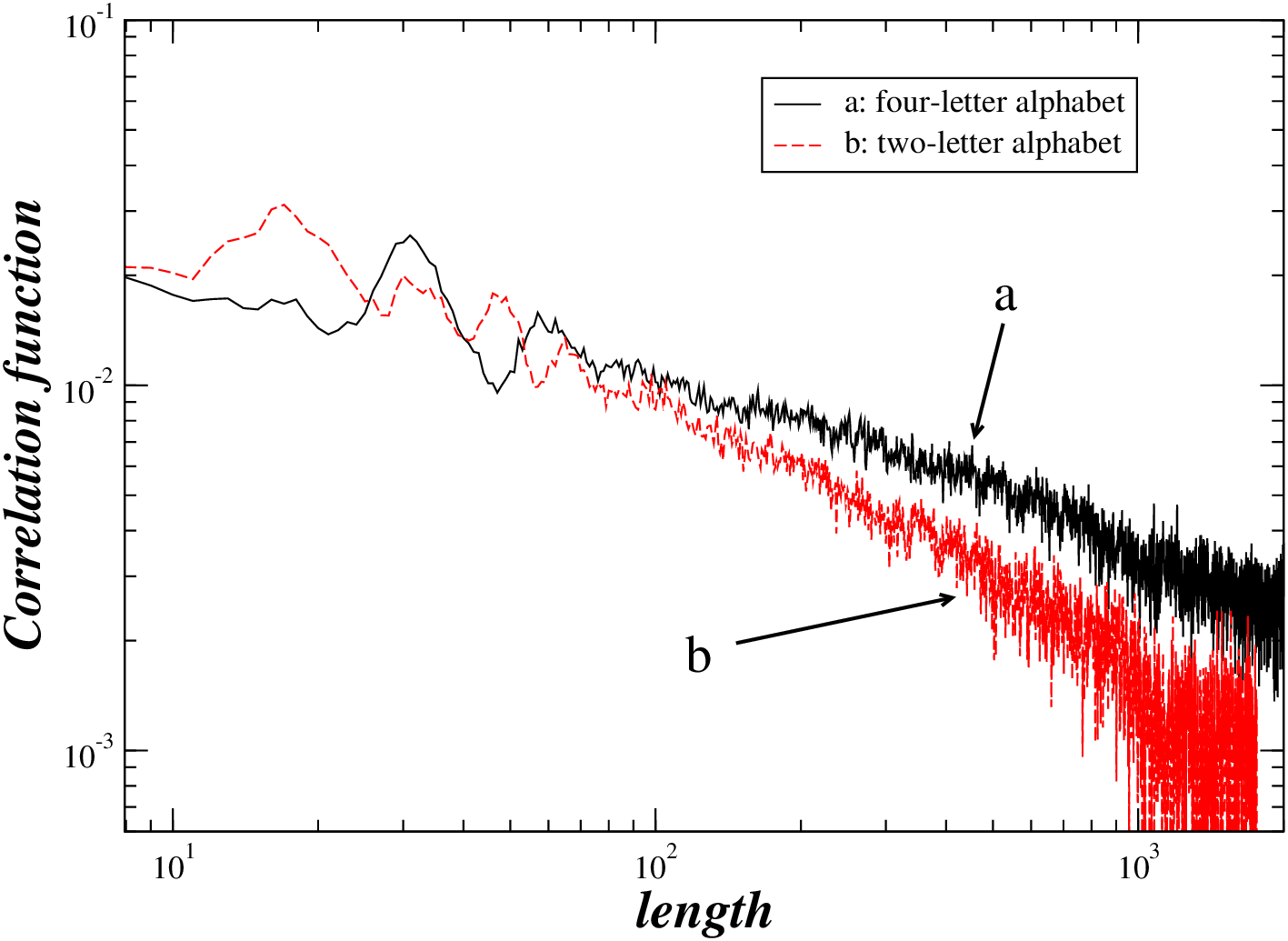} 
\caption{\label{fig:05}{} (Colour online)
Correlation function: four-letter alphabet (curve a) and  two-letter alphabet (curve b).
\hspace{7cm}
}
\end{figure}

 An alternative manifestation of the log-log structure depicted in 
Figs. \ref{fig:04}a,b is that
the individual exit distances display long-range correlations as illustrated in  
Fig. \ref{fig:05}, curves a and b,  for 
the cases of four and two symbols, with power law decaying exponents close to 1/3 and to
1/2, respectively.

\par Table \ref{table:6.1} summarises the mean and variances of the exit distances for the four-
and for the two-symbol cases as compared to the corresponding quantities evaluated
from the Markov chain simulation. We see that while the average 
 values in the two cases are practically indistinguishable
(and equal to the analytic results for a first order Markov chain, $(1-w_{ii})^{-1}$),
the variances associated to the data are larger than the Markov ones. This reflects
the delocalisation of the DNA exit distance distribution in the state space, a property
due among other to the presence of repeats along the sequence.

\begin{table}
\begin{center}
\begin{tabular}{l|c|c|c|c|c|}
\cline{2-6}
 & & \multicolumn{2}{c} {DNA Data} &
\multicolumn{2}{|c|} {1st order Markov} \\
\cline{2-6}
&&$<n>$ & $<\delta n^2>$ & $<n>$ & $<\delta n^2>$ \\ \hline
{\multirow{2}{3.5cm}{Two-letter alphabet} }
 & 1 &2.2725391 & 3.5143790 & 2.2719333 & 2.4828568 \\\cline{2-6} 
                               &2&2.2835724 & 3.7262988 & 2.2843783 & 2.5242081\\\hline\hline
{\multirow{4}{3.5cm}{Four-letter alphabet}}
 & 1 & 1.4796035 & 0.9562388 & 1.4791129 & 0.7088170\\ \cline{2-6}
                                & 2 & 1.3499513 & 0.4599166 & 1.3502789 & 0.4731067 \\ \cline{2-6}
                                & 3 & 1.3498576 & 0.4588480 & 1.3498057 & 0.4724476\\ \cline{2-6}
                                & 4 & 1.4887851 & 0.9832494 & 1.4887587  & 0.7272797\\ \hline
\end{tabular}
\caption{\label{table:6.1} Means and variances of exit distances for the DNA data 
and for a 1st order Markov chain.
\hspace{7cm}
}
\end{center}
\end{table}

\begin{figure}
\includegraphics[clip,width=0.6\textwidth,angle=0]{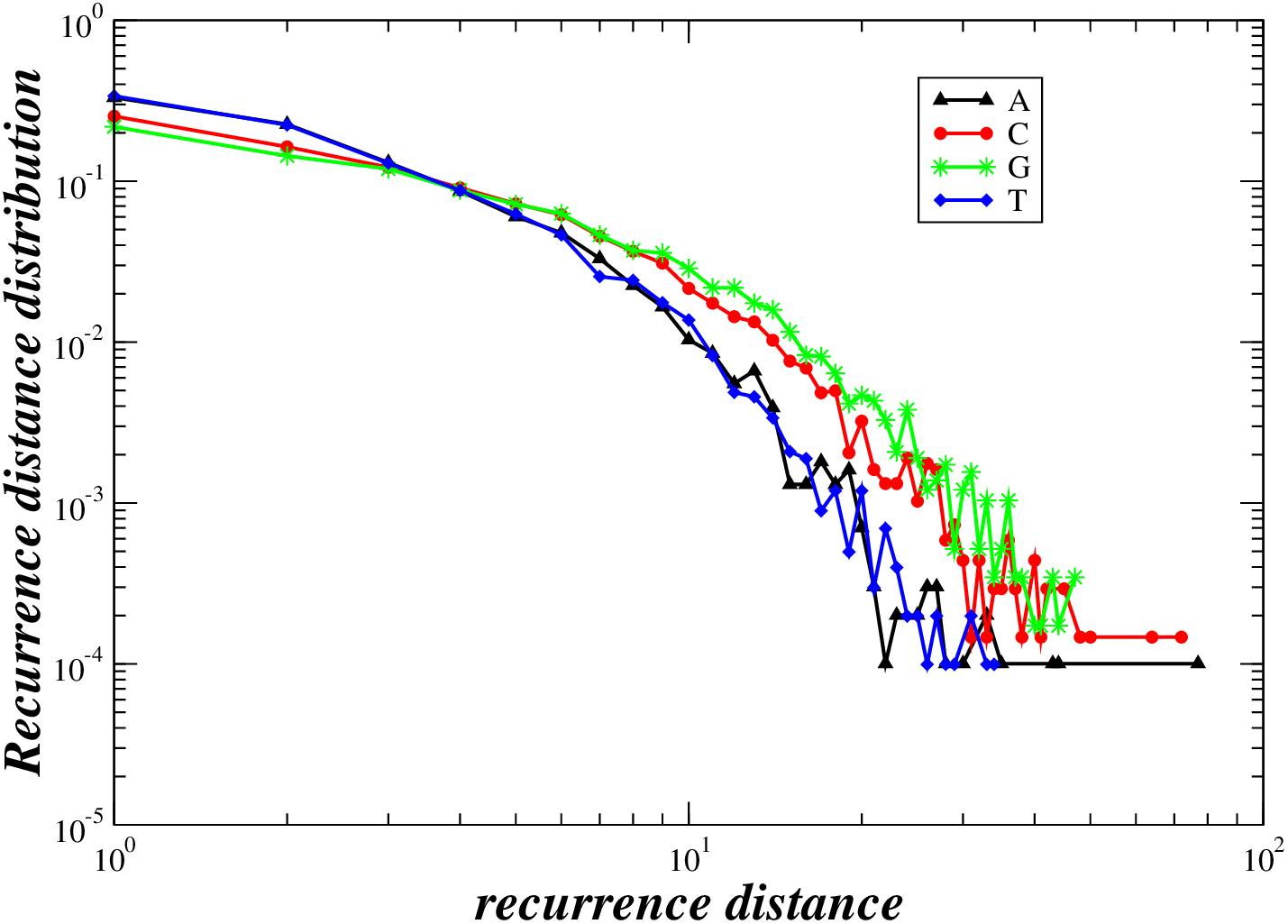}\\

\caption{\label{fig:06}{} (Colour online)
Recurrence distributions of the four symbols.
}
\end{figure}

We turn next to the recurrence distribution $F_{j,n}$. We first observe that in the 
two-letter alphabet the recurrence distribution of one of the two states is fully
determined by the exit distance distribution of the other state. We are thus again in 
the presence of long-ranged distributions and long range correlations of the
individual recurrence distances, 
see Figs. \ref{fig:05} 
and the first two rows of Table \ref{table:6.1}. 
Coming to the four-letter alphabet, as for $q_{j,n}$, practically identical
recurrence distributions for states $A$ and $T$
 and for states $C$ and $G$ are observed, see Fig. \ref{fig:06}. Both distributions
are long ranged with the $A$ and $T$ falling off more slowly
 than for $C$ and $G$ for large
$n$ (and a crossover between the two distributions at $n\sim 5$). 
Furthermore, compared to the corresponding exit distance distributions, they are
more delocalised as illustrated by Table \ref{table:6.2}, to be compared  with
the last four rows of Table \ref{table:6.1}.
\par
As was the case of Table \ref{table:6.1} the averages $<n>$ are very close to those obtained by
simulating a first order Markov chain with a conditional probability matrix provided by 
the data, as well as with the well known analytic result $<n>=1/p_i$.

\begin{table}
\begin{center}
\begin{tabular}{|c|c|c|c|c|}
\hline
 & \multicolumn{2}{c} {DNA Data} &
\multicolumn{2}{|c|} {1st order Markov} \\
\hline
&$<n>$ & $<\delta n^2>$ & $<n>$ & $<\delta n^2>$ \\ \hline
1&3.6221180 & 12.571706 & 3.6272037 & 9.004858 \\ \hline
2&5.1011710 & 29.414993 & 5.1058583 & 20.609692\\ \hline
3& 5.0786543 & 29.119154 & 5.0796456 & 20.368145\\ \hline
4& 3.5896053 & 12.082341 & 3.5923173  & 8.781096\\ \hline
\end{tabular}
\caption{\label{table:6.2} Means and variances of recurrence distances for the DNA data 
and for a 1st order Markov chain.
\hspace{8cm}
}
\end{center}
\end{table}
\vskip 0.5cm

\par 
Closely related to recurrence is the concept of analogs, which finds its origin in the classification
 of atmospheric circulation patterns in meteorology. Translated in the language of the (coarse-grained) 
description of a symbolic sequence the issue is, to what extent there exist persistent patterns in different 
(distant) parts along the sequence, where symbols are found in a given prescribed order with an 
appreciable frequency \cite{nicolisc:1998}.

\begin{figure}
\includegraphics[clip,width=0.44\textwidth,angle=0]{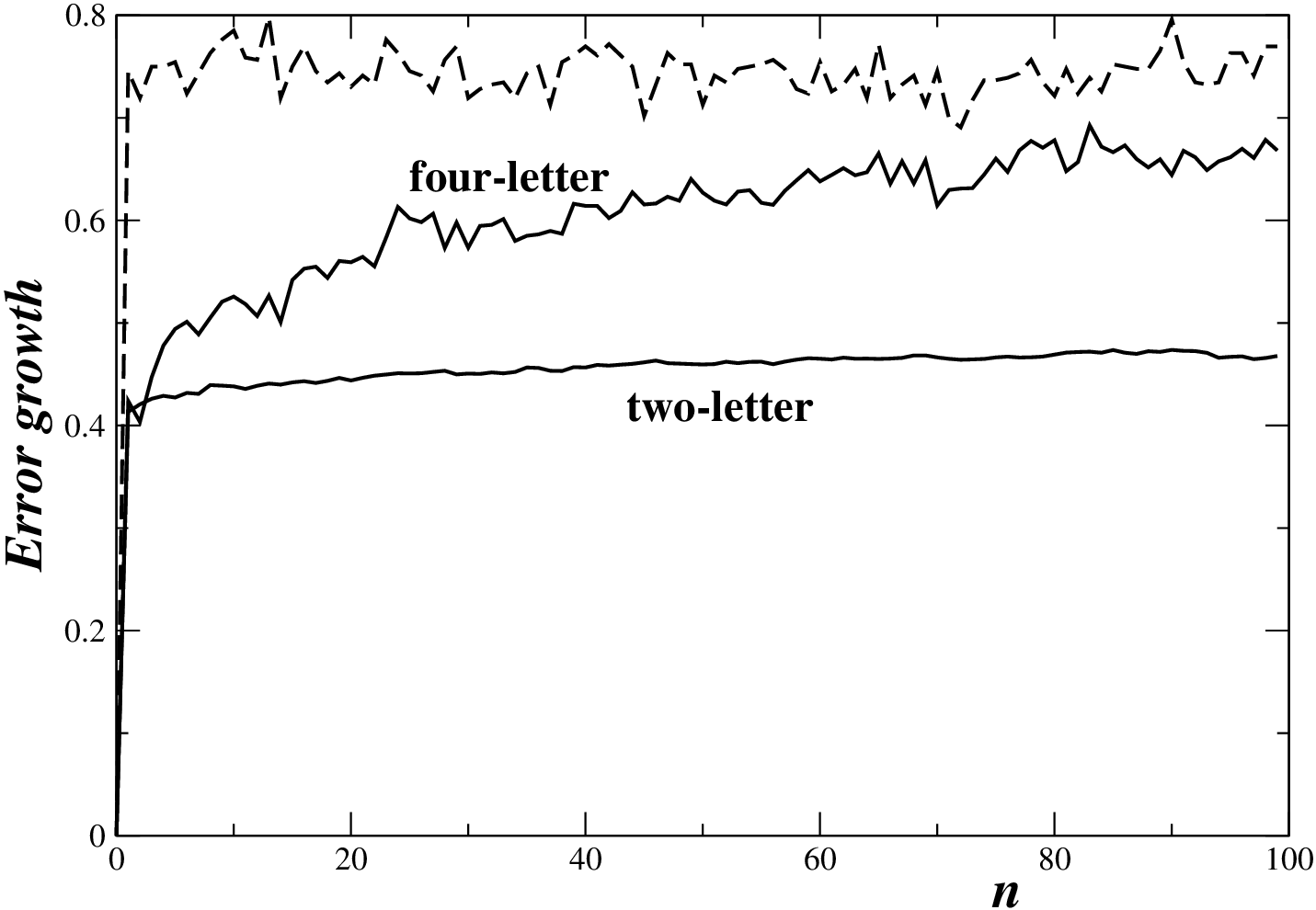}
\includegraphics[clip,width=0.44\textwidth,angle=0]{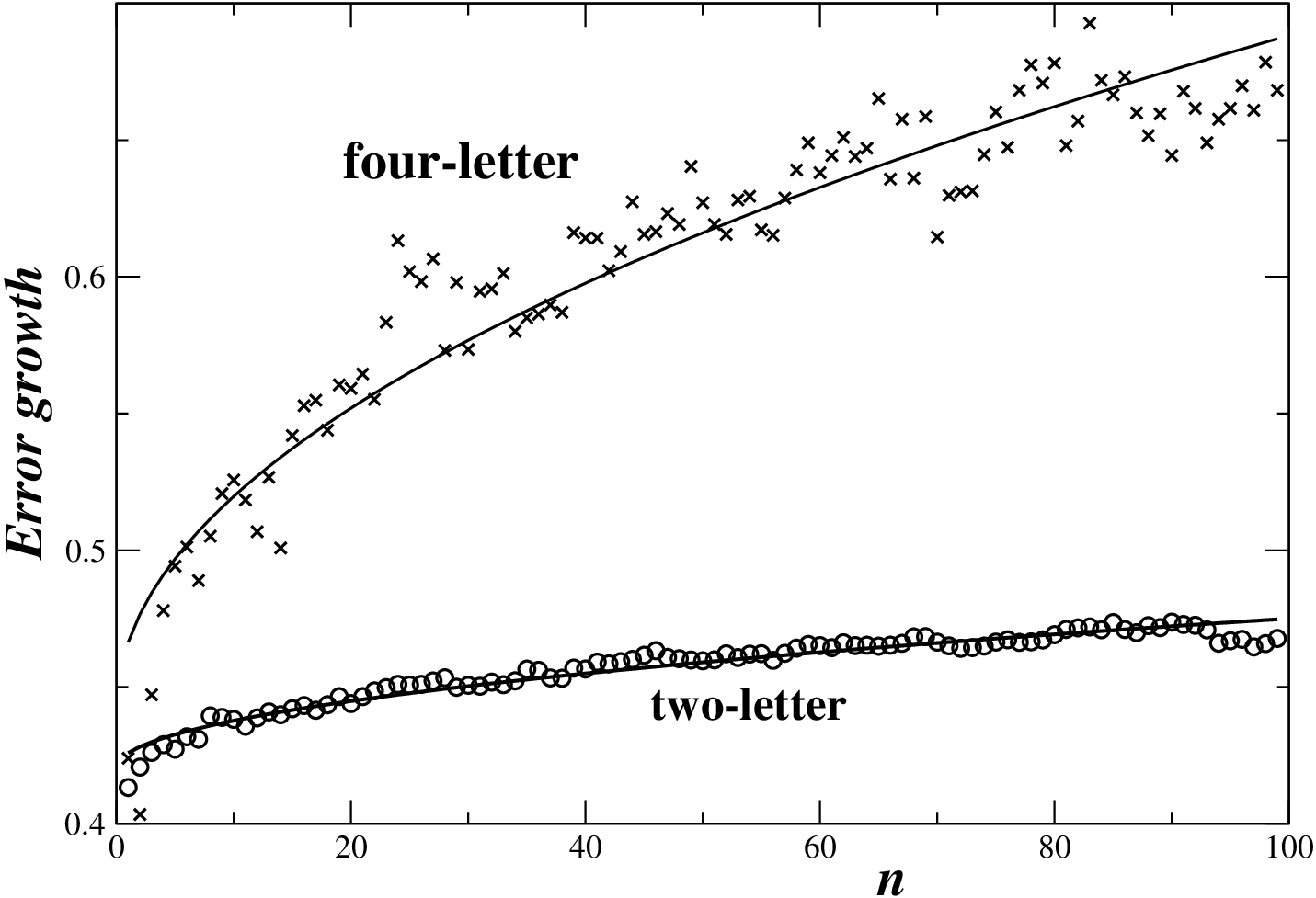}

\caption{\label{fig:07}{}  a) Error growth functions
for the two-letter alphabet and the four-letter alphabet; the 
dashed line represents a random sequence). b) Same as (a) with
solid lines depicting nonlinear fit.}
\end{figure}

To address this question for the DNA symbolic sequence we consider all pairs of n-subsequences along 
the full sequence containing identical symbols in sites $1,\cdots$ up to $m$, and compute the 
``error'' (in the sense of the Hamming distance, see Sec. V) as they start deviating from the 
$(m+1)$-st site and onwards. The result for the two and four-letter alphabets and for 
$n=100, \>\> m=8$ is depicted in Fig. \ref{fig:07}a. The dashed lines in this figure correspond to 
a random sequence. As expected, beyond $n=8$ the symbols in the two members of the pair 
alternate indifferently between being identical (error 0) or being different (error 1), 
entailing that the error attains immediately a saturation value. The situation is very different 
for the DNA data, represented by the solid lines in Fig. \ref{fig:07}a. Here a first stage of abrupt increase 
of the error is followed by a stage of very slow increase toward the saturation level, even though 
this level is not yet attained for $n$ up to 100. This indicates a persistence trend or, alternatively,
 the presence of long-range correlations and is further confirmed by the plot of Fig. \ref{fig:07}b 
suggesting a power law dependence of the error on $n$ prior to the final decay to the saturation 
level with a power  of the order of 0.5.
 This behavior can be viewed as the ``spatial'' analog of the error growth dynamics familiar 
from dynamical systems theory where, after an exponential stage (to be compared with the stage
 of fast growth in Fig. \ref{fig:07}a), one observes a diffusive stage prior to the final stabilization to 
the saturation level. 

\par
Within the set of quantities which probe
the local structure of a sequence, 
the Hurst exponent $H$ expresses the tendency
of the future values of a sequence to persist or increase on average, or to fluctuate between 
small and large values \cite{feder:1988}. In
particular, for the range $0\le H <0.5$ the sequence values tend to alternate,
while for $0.5 < H \le 1$ they tend to persist or increase on average.
The value 0.5 is a border case, where the values are either completely
uncorrelated or their correlations decay exponentially fast to zero.
\par To apply the concept of the Hurst exponent in  DNA sequences (or any symbol sequence in general)
we map the nucleotides (symbols) to numbers. 
For the calculations  we use the
two-letter alphabet $PU-PY$ notation,
to conform with the previous analysis on the exit distance distributions and the mapping 
takes the form:
\begin{align}
PU \to 0 \nonumber 
\\
PY \to 1    
\label{eq25}
\end{align}
Thus the symbol sequence turns into a corresponding numerical sequence,
which carries all the information on the position of symbols. $H$ is then
directly calculated from the numerical series and 
is a significant measure which  expresses the tendency of symbols
to repeat themselves (persistence)  or to alternate (antipersistence) down the sequence. 
\par 
The calculation of the Hurst exponent is based on the computation of the 
range $R(n)$ between the maximum and the minimum cumulative 
values as one advances along the numerical sequence of
size $n$, for various values of $n$. Cumulative values are essential
in the $H$ estimation because they keep track of the  
tendencies along the sequence.
One then needs to rescale $R(n)$ by the standard deviation $S(n)$,
\begin{equation}
S(n)=\sqrt{\frac{1}{n}\sum_{i=1}^n(x_i-<x>)^2}     
\label{eq26}
\end{equation}

\noindent in order to obtain the \textit{rescaled range}. Once the rescaled range
$R(n)/S(n)$ is calculated it is averaged over many sequences (configurations)
of the same length $n$.
 \begin{equation}
E(n)=\langle \frac{R(n)}{S(n)}\rangle_{confs}     
\label{eq27}
\end{equation}
The Hurst exponent is then  defined as
 \begin{equation}
E(n)=c\cdot n^H     
\label{eq28}
\end{equation}
and is computed from the slope of $E(n)$ versus $n$ in a double logarithmic scale.
When the sequence is characterised by fractality, with fractal dimension $D$, it can be
shown that $H=2-D$, where $1< D< 2$.

\begin{figure}
\includegraphics[clip,width=0.6\textwidth,angle=0]{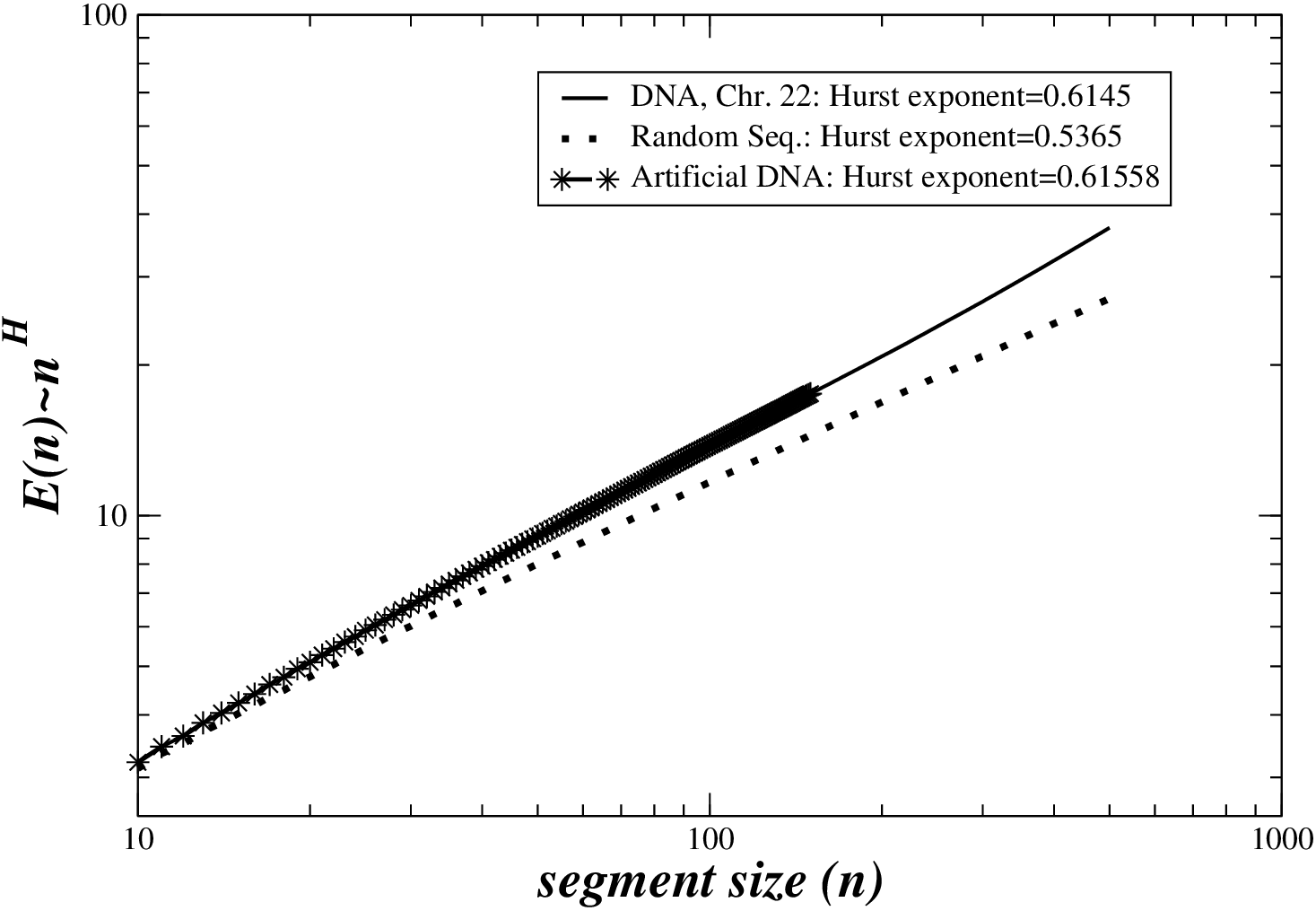}

\caption{\label{fig:08}{} The rescale range $E(n)$ as a function of the sequence size $n$,
for the calculation of the Hurst exponent $H$.\hspace{6cm} }
\end{figure}
\par 
In Fig. \ref{fig:08} the rescaled ranges $E(n)$ are plotted as a function of $n$
for the working contig N\textunderscore 011387 of Chromosome 20 (solid line), 
the random sequence (dotted line) and the model DNA (stars) which will be discussed
in the following section. The value calculated for the Hurst exponent
is $H=0.6145$ and is clearly distinct from that of the random sequence
with the same letter frequency as the data. Calculations of the Hurst
exponent in other 
human contigs give very similar $H$-values. Values of $H>0.5$ indicate
persistence of the same symbols along the sequence, or to put it differently,
clustering of similar nucleotides. This effect is a cause of correlations 
and can reflect the well known existence of poly(A) and poly(T) (in
the complementary chain) motifs in the primary
genomic DNA sequences that give rise to the corresponding poly(A) signals in mRNA
\cite{kalkatawi:2012}. Another source of the clustering of similar nucleotides
is the Alu repeats \cite{hackenberg:2005,price:2004}
 in the human genome which are also known to be associated with 
poly(A) sequences \cite{lustig:1984} . In addition, non-coding gene-poor (desert) regions 
 are known to be rich in A and T inducing clustering of these symbols,
while  CpG rich regions (isochores) are rich in genes and induce 
lower scale clustering \cite{li:2003,oliver:2002,luque:2005,bernaola:2004,arndt:2005}.

\section{A model DNA}
\label{sec:model}
\par DNA, a complex multicomponent structure which has evolved during  billions
of years in close contact with an ever-changing environment, cannot be described or
constructed based on  a closed
functional expression with a limited number of parameters.
 Statistical constructive methods or methods based on chaotic dynamics
 have been used since the early
1990 to create long nucleotide sequences with statistical properties mimicking those
of specific DNA molecules
\cite{stanley:1992,almirantis:2001,gutierez:2001,herzel:1995,provata:1999,allegrini:1995,beck-prov:2011}. 
All these attempts predict well some of the sequence properties  but they fail in others
and one needs to add an increasing number of parameters to probe into the structure's
local details, even from the statistical point of view.
In the present section we propose a novel, 
global statistical construction method based on the exit distance
distribution of the DNA sequence described in Sec. \ref{sec:exit-distances}, to
generate two- and four-letter sequences with statistical properties
basically identical with the ones of the original DNA data.
\par The construction method is known as ``Monte Carlo
rejection sampling'', or simply rejection sampling, and dates back to J. von Neumann. 
Having calculated the exit distance distributions for the segments of all symbols
in the natural DNA sequence, we use the rejection sampling method to 
create a statistically equivalent series. For simplicity, the method
 is described in the two-letter alphabet and is easily extendable to the four-letter one.
\begin{enumerate}
\item Define first the initial symbol as a $PU$ or $PY$, either randomly 
        or as dictated by the contig sequence.
\item Select an integer random number between $\left[ 1,N_{PU}^{max}\right] $ or
 $\left[ 1,N_{PY}^{max}\right] $
depending on whether a $PU$ or a $PY$ segment is to be created. [$N_{PU}^{max}$ and
 $N_{PY}^{max}$ are the maximum numbers of juxtaposed $PU$'s or $PY$'s which have been
observed in the natural contig]. Call the selected number $n$.
\item Chose a second random number $r\in \left[ 0, 1\right]$ and compare it to the value
      of the exit distance distribution $q_{PU,n}$ or $q_{PY,n}$ depending on the
      current state on the chain.
\item If  $r\le q_{PU,n}$ (or $r\le q_{PY,n}$) then the sequence is extended by $n$
      units of $PU$ (or $PY$).
\item The algorithm returns to step 2 in order to make alternating additions of 
       $PU$ and $PY$ clusters.
\item The algorithm stops when the size of the artificially constructed sequence
        is equal to the size of the natural DNA contig.
\end{enumerate}
The artificial sequences created with the rejection sampling method are constructed to respect 
perfectly the 
exit  distance distributions of the natural sequence. They possess all other statistical
 properties of the natural sequence as well, since the knowledge of the exit distance distribution
alone is sufficient to carry out the construction. In particular, exit from a given state implies
automatically entrance to the complementary state in the two-letter alphabet.
The situation is different in the four-letter case.
Here, one more assumption needs to be made regarding the
alternation between the four symbols. Our procedure is based on the transition
probabilities $w_{ij}$ between the different letters as were presented in Table \ref{table:2.1}
and implies thus the assumption that higher order transition probabilities are not
accounted for at this stage.

\begin{figure}
\includegraphics[clip,width=0.45\textwidth,angle=0]{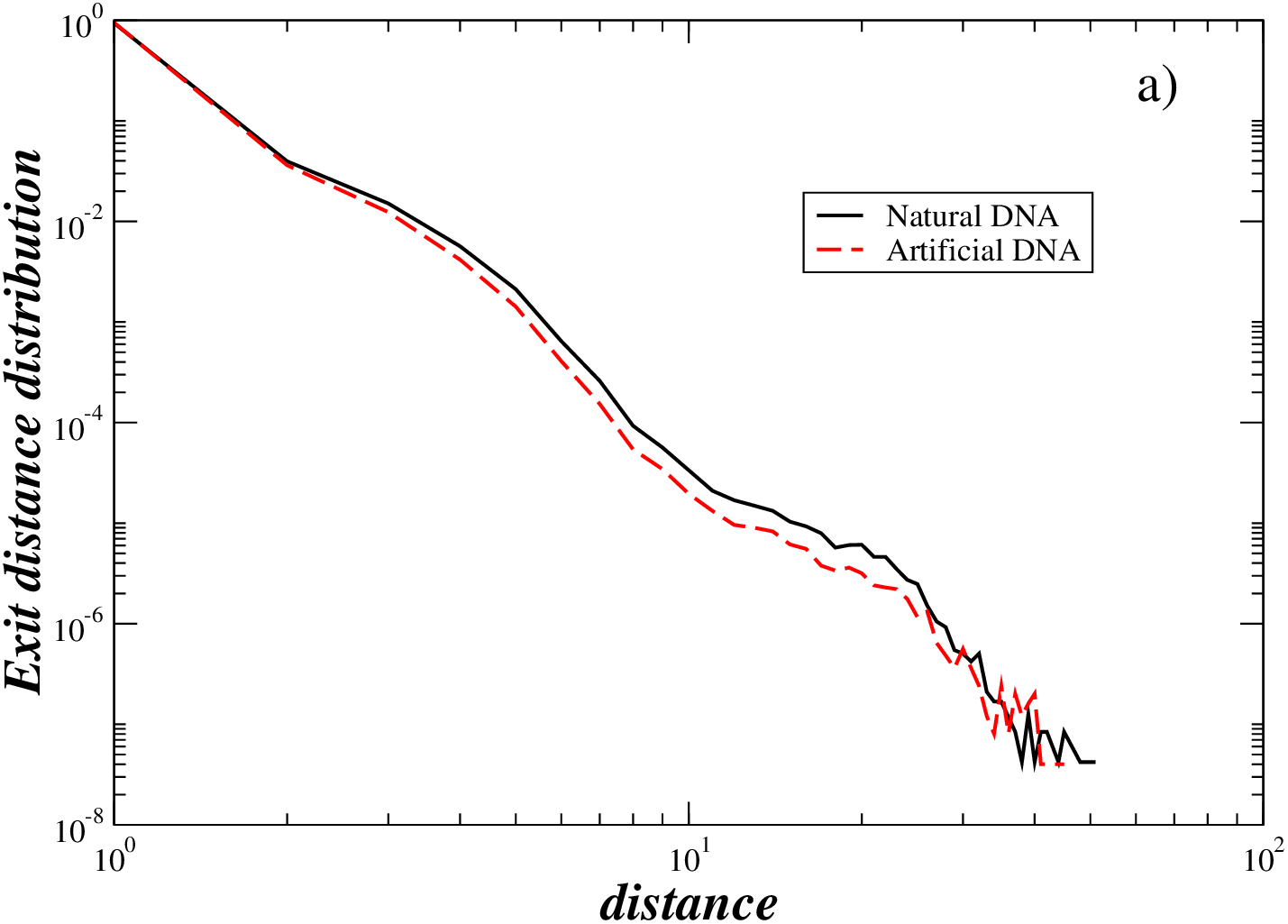} 
\includegraphics[clip,width=0.45\textwidth,angle=0]{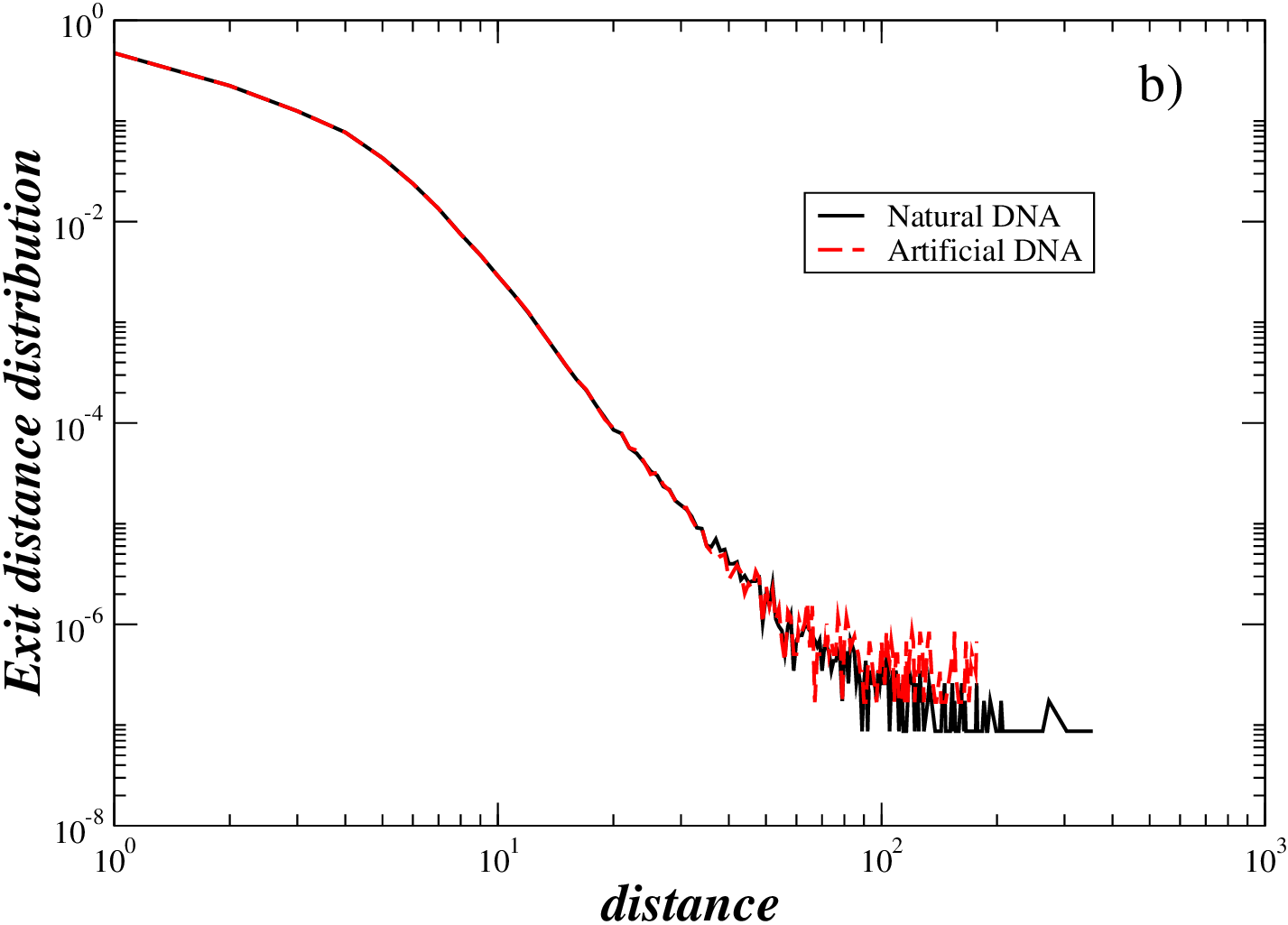} 
\caption{\label{fig:09}{} (Colour online) Exit distance 
distributions for natural (solid black lines) and artificial DNA (red dashed line)  
: a) four-letter alphabet, $A$-symbol; b) two-letter alphabet, $PU$-symbol. 
\hspace{14cm}  }
\end{figure}

In Fig. \ref{fig:09} the exit distance distributions for the $A$ symbols (four-letter alphabet)
and the $PU$ coarse grained symbol (two-letter alphabet) are shown, both for the
original and the artificial model-based DNA sequence; similar plots are obtained for
the other symbols. As one can see, the 
distributions are almost identical in the case of the two-letter alphabet, 
Fig. \ref{fig:09}b,
with small differences in the tails of the distribution attributed to the finite size
of the sequences. The differences are non-trivial in the case of the four-letter alphabet,
Fig. \ref{fig:09}a,
and this is attributed to the use of $w_{ij}$'s which account only for the pair correlations,
while for the juxtaposition of segments of size $n$ higher order correlations 
(of range up to $n$) need to be taken into
account.

\par In Sec. \ref{sec:entropy}, Fig. \ref{fig:02}, 
the information transfer $I$ between a sequence and its shifts
 was shown, both for the original and for the model-generated sequence. Both sequences
show the same degree of information transfer in first and second neighbour positions. 
However, for more distant positions the information transfer in the model-generated sequence
undergoes an abrupt decay as compared with the natural DNA sequence where the information
transfer persists for hundreds of units. This difference reflects the
functional role of natural DNA sequence, as opposed to the statistical character of the
artificial DNA. The nucleotides in a natural sequence need to control the information 
about neighbouring positions, since what dictates their functionality is
their precise (not statistical)
juxtaposition. In particular, information flow in
decades of bps relates to the turn of the helix, while information flow in a few 
hundreds of bps is plausible, since these are typical sizes for coding regions and for repetitive
elements.
\par In a similar vein, the analog analysis shows that the error values as obtained from the
model lie closer to the saturation level than those obtained from the natural DNA as 
depicted in Figs. \ref{fig:07}. Interestingly, the Hamming and modified Hamming distances 
between the natural and the model sequences equal to 0.7444 and 0.6222, 
respectively and are close to the values 
 associated with distances between random sequences.
\par Finally, in Sec. \ref{sec:exit-distances}, Fig. \ref{fig:08},
the Hurst exponent $H$ is depicted both for  the natural (solid line) and for the 
model sequence (stars). By its nature $H$ is a nonlinear measure which takes into account all
orders of correlations since it deals simultaneously with all segment sizes. As can be
seen from the figure the curves $E(n)$ for the natural and the model sequence are practically
indistinguishable and the values of $H$ are very close.

\section{Conclusions}
\label{sec:concl}

In this study
 the structure of global human chromosomal sequences has been analyzed using 
ideas and tools from nonlinear dynamics, information and complexity theories 
and nonequilibrium statistical mechanics. We have shown that in the four-letter 
alphabet the sequence exhibits spatial asymmetry and suggested on these grounds 
that the DNA molecule can be viewed as an out-of-equilibrium structure. 
We have established a connection between these properties and the generation 
of long-range correlations and the processing of information along the sequence, 
using a series of entropy-like quantities. We have introduced the exit and 
recurrence distance distributions, two new indicators of the complexity underlying 
the sequence, whose evaluation revealed a number of interesting features of 
its global structure. Finally we have designed an algorithm generating sequences 
that share the statistical properties of natural DNA, local as well as global, 
on the sole basis of the exit distance distribution. The results reported 
pertain to human chromosome 20. Other chromosomes (10, 14 and 22) have been tested 
and shown to lead to similar conclusions.

The approach initiated in this work opens some interesting and worth-exploring 
perspectives. A first line of approach would be to apply the ideas of asymmetry, 
irreversibility and information processing considered here in a global perspective 
to particular DNA building blocks such as coding DNA, non-coding DNA, repeats, etc. 
Another case to consider are higher eukaryotes, whose genomes share with human genome 
the existence of genes separated by long non-coding regions containing a high 
concentration of repeats. Similarly, in the spirit of comparative genomics, 
it would be interesting to apply the ideas developed here on organisms with 
intrinsically different genomic structure such as prokaryotes versus eukaryotes. 
\par
Finally, a quantitative comparison between the 
local and global statistical properties of the human genome derived in this work and 
those of the genome of higher mammals and especially of primates could lead to  
striking evolutionary insights.

\textbf{Acknowledgements:} A. P. acknowledges financial support from the National Center for
Scientific Research ``Demokritos'' for a Sabbatical visit to the Universit\'e Libre 
de Bruxelles and the European Science Foundation programme ``Exploring the Physics of 
Small Devices'' for the scientific exchange grant EPSD-4308
with the same university. We also acknowledge an interesting discussion with
 Professor M. G. Velarde.

\end{document}